\documentclass[iop]{emulateapj-rtx4} 
\shortauthors{Sekanina}
\shorttitle{Historical Controversy and Future of Bright Sungrazing Comets} 
\slugcomment{Version \today }

\begin{document}
\title{Bright Sungrazing Comets in a Great Historical Controversy
 and\\Prospects for Their Appearance in the Near Future\\[-1.6cm]}
\author{Zdenek Sekanina}
\affil{La Canada Flintridge, California 91011, U.S.A.; {\sl ZdenSek@gmail.com}}

\begin{abstract} 
Until the second half of the 19th century, two or more brief appearances
of bright comets,~such~as~the ones in 1668 and 1702,\,alike in aspect and
motion,\,seen with a tail near the Sun,\,were
almost universally believed to be
periodic returns of a single object.  It is likely that the exceptional
story of Halley's~comet was~the compelling precedent for this school of
thought.  Application to sungrazers was discredited~by the~observed
%
%
%
fragmentation of the nucleus of the giant sungrazer of 1882 shortly after
perihelion.  Generally, separations and orbital periods of the Kreutz
comets are known to be governed in such~events~by the~solar~tidal
force, while the range in the longitude~of~the~nodal~line~is~linked~to~the
pyramidal architecture caused by nontidal, cascading fragmentation
along the entire orbit and described by an updated contact-binary model.
Perception of the sungrazer system was changed dramatically by coronagraphic
imaging from space, which led to discovery of up to ten populations
of~dwarf~comets.  Past~fragmentation patterns have been used to
tentatively predict the arrivals~of~two~bright~Kreutz~sungrazers~---
a~Population~II member around 2027 (and before 2040) and a Population~I
member~around~2050.{\vspace{-0.05cm}}
\end{abstract}
\keywords{individual comets: comet of 372 BC, 363, 1041, X/1106 C1, 1138,
 C/1668~E1, C/1689\,X1, C/1695\,U1, X/1702\,D1, C/1843\,D1, C/1880\,C1,
 C/1882\,R1, C/1887\,B1, C/1963\,R1, C/1965\,S1, C/1970\,K1, C/2011\,W3,
 C/2024\,S1; methods: data analysis\vspace{-0.1cm}}

\section{Introduction} 
Massive sungrazing comets, readily detected with the naked eye, are among
the brightest cometary objects ever observed, always displaying a prominent
tail for days or weeks after perihelion.  However, because of the orbital
geometry, the level of spectacle for the Kreutz sungrazers depends on the
time of the year the object arrives at perihelion, the most favorable
months being January through mid-March for evening displays, but September
through October for morning displays.  The southern hemisphere is decidedly
preferred.  Because of their unknown (or poorly known) orbits, the potential
candidates that had arrived before 1843 could not be identified with certainty
and are in the following referred to as possible (or likely, in more hopeful
cases) Kreutz sungrazers.

The naked-eye objects considered today to be the definite Kreutz sungrazers
are, chronologically, the Great March Comet of 1843 (C/1843~D1), the Great
Southern Comet of 1880 (C/1880~C1), the Great September Comet of 1882
(C/1882~R1), the Great Southern Comet of 1887 (C/1887~B1), Pereyra
(C/1963~R1), Ikeya-Seki (C/1965~S1),{\vspace{-0.04cm}} White-Ortiz-Bolelli
(C/1970~K1), and Lovejoy (C/2011~W3).\footnote{I do not include C/1882~K1
(Tewfik), a dwarf sungrazer that apparently was seen with the naked eye
only during the total solar eclipse on 1882 May~17, several hours before
passing perihelion.}  Among historical comets, selected possible Kreutz
sungrazers are Aristotle's comet of 372~BC, X/1106~C1, a Korean comet of
1041 and a Chinese comet of 1138 [Nos.\ 371 or 372 and 403, respectively,
in Ho's (1962) catalogue], C/1668~E1, C/1689~X1, C/1695~U1, and X/1702~D1.
In the context of the contact-binary model of the Kreutz system (Sekanina
2021), a group of daylight comets late in AD~363, noted briefly by the
Greek historian Ammianus Marcellinus, fills in a major gap in an early
phase of evolution.

%

\section{The History of Investigation of\\Bright Sungrazers}
It is not widely known that the history of scientific~debate on the nature
and whereabouts of the Kreutz sungrazers is at least two centuries older
than their name.  Kreutz (1901) knew that~the~bright~comets~of~the~years
1668 and 1702 were thought to be the same object.  It was in fact Schumacher
(1843), the founder of the {\it Astro\-nomische Nachrichten\/}, who
introduced the story that started the controversy on the occasion of
the appearance of the Great March Comet~of~1843~(C/1843~D1).
Schumacher reported that he was alerted by Mr.\ Cooper from
Nice\footnote{Obviously E.\ J.\ Cooper, the Irish astronomer, who in the
1840s was traveling several European countries, including France.} that
in his book {\it L'usage des Globes\/}~N.~Bion noted that Cassini (1702)
proposed the identity of the comets~of~1702 (X/1702~D1) and 1668
(C/1668~E1), thus implying an orbital period of 34~years, while the
object's existence stretched over $\sim$2000~years (back to 372~BC).
Accordingly, three sungrazers figuring in the majority of modern
scenarios for the Kreutz system were in the focus of scientific
interest more than 300~years~ago.

Pointing out that Cooper thought that the 1843 sungrazer was identical with
both the 1668 and 1702 comets, Schumacher offered a few critical comments
of his own.  He noted that if the three comets were the same returning
object, the 1843 appatition would be the fourth arrival to perihelion after
1702 and the mean orbital period would then be 35$\frac{1}{4}$~years,
not 34~years.  He also deemed it difficult to explain why the bright
and long tail was overlooked in each of the three intervening returns
between 1702 and 1843.  Schumacher suggested that the 1843 comet might
be identical with only one of the two earlier comets and that one should
take pains to carefully check either of the two options.

Although successive appearances of the same object~{\it or\/} surviving
fragments of a common parent are two a priori equally plausible options
of contemplating the relationship between two or more bright comets near
the Sun, alike in aspect and motion, it appears that the story of Halley's
comet was a compelling precedent to overwhelmingly prefer the first option,
%
%
even if blemished by adverse implications.\footnote{Notable pre-Kreutz
exceptions in the 19th century, M.\ Hoek and D.\ Kirkwood, are singled out
at the end of this section.}
In the just described case this happened to both Cassini\footnote{Halley's
work ``{\it A Synopsis of the Astronomy of Comets,\/}'' in which the identity
of the comets of 1456, 1531, 1607, and 1682 was proposed and return in 1758
predicted, was published in 1705, three years {\it after\/} Cassini's remark
on the comets of 1668 and 1702.  However, Halley expressed his suspicion on
the 1456--1682 comets already in a 1695 letter to Newton.  Cassini was in
contact with Halley from 1680 on (if not earlier), and it is very likely
that he was aware of Halley's impending findings long before their
publication and they were affecting his scientific thinking about comets.}
and Cooper.  In another example, reporting that the motion of the sungrazer
C/1689~X1 was closely fitted by the orbit of C/1843~D1, Kendall (1843)
likewise affirmed that ``[{\it i\/}]{\it t has never happened \ldots that
two comets have appeared with elements agreeing so well,
without being found in the~end~to be the same.}''  Another mix of these
comets, also including one from 1406, was favored as returns of a
single object by Clausen (1843).\,\,\,\,\,

The story repeated itself later in the 19th century, when the pair
of C/1843~D1 and C/1880~C1,~whose orbital similarity was exceptional, became
a favorite~target; this time the suspected orbital period was obviously
37~years (Section~3.12).  And, as I remarked elsewhere (Sekanina 2022a),
the hypothesis was also tried on C/1843~D1 and C/1882~R1, even though
their nodal lines differed by almost 20$^\circ$ and perihelion
distances by 50~percent.  Some used the 1880 and 1882 sungrazers to
propose an absurd orbital period of 2.6~years!

The speculations about recurring returns of a single object were hard to
refute in early studies of sungrazers, when the orbital period could not
be determined from meager astrometric information available.  But when
Hubbard (1851, 1852) completed his comprehensive orbital investigation
of C/1843~D1 and obtained an orbital period of 533 or 803~years, with a
standard error of less than $\pm$150~years, continued adherence to much
shorter periods became increasingly futile.  The final blow came with the
observations, starting shortly after perihelion, of the multiple nucleus
of C/1882~R1, products of {\it fragmentation\/} of the comet's original
mass at, or very close to, perihelion (Section~3.13).  As if not enough,
Kreutz (1891) subsequently showed that the preperihelion and post-perihelion
observations of this sungrazer could easily be linked, ruling out any
detectable effect by the solar corona on the comet's orbital motion,
such as rotating the line of nodes or manipulating the perihelion
distance.  Fragmentation won!

Hoek (1865a, 1865b, 1866) was obsessed with ``systems of comets''.  The
objective of his research, based on the premise of an interstellar origin
of comets (his motto being ``{\it comets come to us from some star or
other\/}''), was to demonstrate that ``[{\it t\/}]{\it here are systems
of comets in space that are broken up by the attraction of our Sun
\ldots\/}''
%
%
Hoek found that the apsidal lines of a fairly large number
of comets had a tendency to be aligned, but the significance of this
alignment has never been satisfactorily explained.

On the other hand, Kirkwood's (1880) comments were directly responsive
to the appearance of the sungrazer C/1880~C1 and the considerable degree
of similarity between its orbit and the orbit of C/1843~D1.  Kirkwood
clearly preferred a breakup of the parent (which he speculated may
have been Aristotle's comet) at perihelion over a returning object
and suggested that the gap of 37 years between their appearances was an
accumulated effect of the rate of their separation over the 22~centuries.
While the experience with the nuclear splitting~of~C/1882~R1 has indicated
that Kirkwood underestimated the effect by one order of magnitude, the
gist of his idea was correct.\footnote{In the contact-binary model,
this same thought is being used, except that the fragments are
C/1843~D1 and C/1882~R1 and in order to get the numbers right, the
breakup of Aristotle's comet had to take place not at perihelion in
372~BC but near the following aphelion.  Orbital integration supports
the idea that the observed 39.5-year long gap between the two brilliant
sungrazers of the 19th century was an accumulated effect of the rate
of their separation.}  As for an ultimate test, Kirkwood just mused:\
``{\it Should\/} [{\it the sungrazer of 1880\/}] {\it return about 1916
or 1917 we may conclude that it is identical with that of 1843; if not,
the hypothesis of a common origin at a remote epoch may be regarded as
probable.\/}''  Of course, no such comet appeared.

\section{Comments on the Kreutz Sungrazers\\Involved in the
 Controversy} 
In the following I comment on each bright Kreutz sungrazer involved in
the controversy, including peculiar features of the orbit and its
determination.  The problem with a review of this kind is that the
logical order to follow is chronological, but the knowledge has
been building up in the opposite direction, starting with the current
status and progressing back into the past, based on gradually accumulating
evidence linked with (plausible) conjecture where necessary.  Given
this contradiction, I begin with the key milestones in the evolutionary
path of the Kreutz system.{\vspace{-0.1cm}}

\subsection{Fragmentation Process, Evolution, and Hierarchy\\of the
 Kreutz System} 
%
I noted in Section 2 that it was not until the very~end~of the 19th
century when it became generally accepted --- thanks largely to the
work by Kreutz (1888, 1891, 1901) --- that the observed sungrazers
were independent products of a massive primeval comet,~not returns of
the same object.  This amounted to admission that the Kreutz system's
evolution was being governed by fragmentation --- Kreutz used the
word {\it partition\/} or\,{\it division\/} ($\!${\it Teilung\/} in
today's German; he spelled it\,{\it Theilung\/}, common in the German
of the 19th century).

The next question was what force was it that caused the sungrazers to
fragment.  Before the invasion of the sungrazers in the 1880s, the only
comet whose duplicity was extensively{\vspace{-0.03cm}} observed was
3P/Biela (e.g., Maury 1846),\footnote{Another split comet was Liais
(C/1860~D1),~which however was poorly observed.  It had fragmented
in mid-September~1859 at a heliocentric distance of 2.5~AU (Sekanina
1980)~and~the~orbits~of both the primary and companion were
derived~by~Pech\"{u}le\,(1868).} which had split in 1840 at a distance
of 3.6~AU from the Sun (Sekanina 1980).  When the Great September Comet of
1882 broke up into several fragments in close proximity of perihelion and
Kreutz (1891) computed their orbits, one would expect that he recognized
the~undeniable influence of the tidal force{\vspace{-0.04cm}} ({\it
Gezeitenkraft\/} in German) exerted by the Sun.\footnote{The
Roche limit was long known in the 1880s
(Roche 1849, 1850, 1851).  The extreme physical conditions in the solar
corona may be a contributing factor, but fragmentation of comets in close
proximity of Jupiter (16P/Brooks, D/1993~F2 Shoemaker-Levy) shows that
the tidal force dominates.{\vspace{-0.1cm}}}  Instead, however, he rather
vaguely contemplated (on pp.\,53--61) that the partition was caused by a
``{\it disturbing force that developed in the interior of the comet upon
its approach to the Sun, which must have affected the individual cometary
particles differently\,}'' and that based on the contemporaneous knowledge
of the physical constitution of cometary nuclei the question of the nature
of the disturbing force could not be answered.  About three quarters of a
century later Marsden (1967) pointed out that ``{\it Kreutz presumed that
the members of the group had been formed when some primordial comet had
split up at perihelion\/}'', but the tidal nature of sungrazers'
fragmentation very close to the Sun was mentioned for the first time
in his follow-up paper (Marsden 1989).

Marsden's era of sungrazer studies brought a new major issue into the
forefront of interest.  As briefly noted in Section~2, the orbits of the
members of the Kreutz group (or system) are by no means identical.
When it became clear, thanks to Kreutz, that the solar corona had no
detectable effect on the motions of sungrazers and that the orbital
differences observed among them could not be explained by some sort of
a friction during the perihelion passage, the problem of the origin of
the orbital disparities demanded explanation.

Marsden (1967) recognized that there were {\it two subgroups\/} of the
Kreutz sungrazers that he referred to as I and II.  He made a great
effort to provide an insight into the nature of this problem, while
adhering to Kreutz's point of view about fragmentation.  Even though
he applied this hypothesis with stunning success to show that the
Great September Comet of 1882 and Ikeya-Seki from 1965 separated
from one another at their previous passage through perihelion in
the 12th century, complications arising from the constraint on
fragmentation limited to perihelion were about to haunt him,
stretching his extraordinary problem-solving capacity to a maximum.

Perihelion fragmentation was responsible for large variations in the
sungrazers' orbital periods (up to a few centuries) but for hardly any
variations in the angular elements (the longitude of the ascending node
and inclination, in particular) and in the perihelion distance.  Yet, it
was a large spread in these elements that separated the two subgroups.
The indirect planetary perturbations, which depended on the positions of
the planets, Jupiter in particular, at the time of the comet's perihelion,
were the only means left to explain the spread.  Their integrated effect
would have to continue to increase systematically with time, so that at
some point in the distant past it was nil.  Marsden (1967) suggested that
the age of the Kreutz group was \mbox{10--20}~revolutions about the Sun,
an estimate that could be brought down a little in the barycentric
coordinates (Marsden 1989).  However, when Sekanina \& Kracht (2022)
actually did integrate the orbits of the Great March Comet of 1843
(Subgroup~I) and the Great September Comet of 1882 (Subgroup~II) over
two revolutions, from the 19th century back to the 4th century, it
turned out that the differences between the two comets actually {\it
decreased\/} in the absolute value with time in all four elements,
by 2$^\circ\!$.6 per revolution in the longitude of the nodal line
and by 0.11~$R_\odot$ per revolution in the perihelion distance!  The
perturbations happened to work in the~opposite direction than
Marsden needed and could not explain the gap between the subgroups.
 
Astonishingly, Marsden had the solution to the problem at his fingertips
but his conviction that the sungrazers were splitting only at perihelion
was steadfast.  In his 1967 paper he wrote that ``[{\it o}]{\it
ne could always explain the differences between the orbits of the
subgroups by a separation at\/} [{\it a\/}] {\it velocity\/} [{\it of
a few meters per second\,}] {\it near aphelion.  Although most of the
comets observed to split have done so for no obvious reason, one does
really require an explanation when the velocity of separation is some
20\% of the velocity of the comet itself!\/}''  

These statements are certainly correct, except that I fail to comprehend
why, for example,{\vspace{-0.01cm}} a separation velocity of 4~m~s$^{-1}$
needs to be explained{\vspace{-0.01cm}} when the orbital velocity is
20~m~s$^{-1}$ (in a presumed case of breakup at aphelion of a Kreutz
sungrazer) but{\vspace{-0.03cm}} does not need to be explained when
the orbital velocity is 20~km~s$^{-1}$.  The velocity of separation is a
function of the fragmentation process and is independent of the orbital
motion.  The physics of a breakup is oblivious to, and has nothing in
common with, the velocity about the Sun.

The constraint that sungrazers fragment only at peri\-helion was at the root
of another problem that complicated Marsden's (1989) interpretation of
the temporal distribution of the dwarf comets detected by the coronagraphs
on board the Solwind and Solar Maximum Mission satellites.  He noted that
some of these objects were arriving in pairs, in extreme cases only about
12~days apart.  If such two fragments had separated from one another at
perihelion some eight centuries before, Marsden calculated that their
relative nongravitational acceleration responsible for a 12~day gap would
have amounted to only \mbox{$4 \:\!\!\times \! 10^{-10}$}\,the solar
gravitational acceleration, about five orders of magnitude lower than the
lowest nongravitational accelerations derived for companions of the split
comets, implying that it would take years for the separation velocity to
reach the velocity of escape.  This would not make much sense and the
obvious solution was to accept that the separation took place not at
perihelion but much more recently.  Indeed, modeling suggests for example
that in a sungrazing orbit with a period of 900~years two fragments arrive
at perihelion 12~days apart when they separated with a relative radial
velocity~of~1~m~s$^{-1}$ at a point of the orbit approximately 130~AU from
the Sun some 150~years before arriving at the following perihelion (and thus
$\sim$750~years after the previous perihelion).  There is an infinite number
of scenarios that would produce the same gap of 12~days, depending on the
magnitude and direction of the separation velocity, the magnitude of the
nongravitational acceleration, and the orbital position at separation.  The
dimensions of the orbit are of course also important.  The lesson one learns
is that the {\it sungrazers break up at any point of the~\mbox{orbit}\/}.
I arrived at this same conclusion (Sekanina 2000) from independent evidence
on the episodic distribution, on much shorter time scales (as short as a
small fraction of a day), of the dwarf Kreutz sungrazers observed in
large numbers in the images taken with the C2 and C3 coronagraphs on
board the Solar and Heliospheric Observatory (SOHO).  Evidence
on 3P/Biela and other comets observed more recently shows that they
do indeed fragment anywhere along the orbit.

The chance of fragmentation events occurring along the entire orbit
has a major impact on the evolution of the Kreutz system.  Observations
indirectly confirm that fragmentation progresses in {\it cascading\/}
fashion, whereby each fragment of a \mbox{$k$-th} generation produces
a number of smaller fragments of the \mbox{$(k\!+\!1)$-st}
generation, each of which in turn produces a number of still smaller
fragments of the \mbox{$(k\!+\!2)$-nd} generation, etc.  The process
eventually generates fragments that are too small to survive perihelion.
This is the case of all so-called dwarf sungrazers imaged by the
coronagraphs on board SOHO and the other spacecraft whose primary
purpose has been the research of the Sun.  We are firsthand witnesses
to the terminal phase of evolution --- the complete disintegration ---
experienced by a large number of members of the Kreutz system.

In a broader sense, fragmentation of sungrazers around the orbit means
{\it much higher fragmentation rates\/} in comparison to those driven
by fragmentation confined~to~perihelion.  This increase implies in turn
{\it shorter lifetimes\/} of fragments and necessitates a {\it tight limit
on the age\/} of the Kreutz system.  Indeed, the contact-binary model
implies that the Kreutz system is currently 2$\frac{1}{2}$ revolutions
about the Sun old, equivalent to approximately two millennia.\,\,\,\,

The recognition of the two subgroups represented in effect an initiation
of the research on the {\it hierarchy\/} of the Kreutz system.  Since
Marsden's pioneering work, large amounts of new information have emerged
leading to considerable progress achieved in the study~of~this~subject.
An early surprise arrived~three~years~after~Marsden's first paper was
published.  The new sungrazer,~White-Ortiz-Bolelli (C/1970~K1), was a
member of neither Subgroup~I nor Subgroup~II; the longitude of its
ascending node was almost exactly 10$^\circ$ smaller than that of
comet~Ikeya-Seki (C/1965~S1), while the perihelion distance was 0.24~$R_\odot$
greater.  The differences relative to Subgroup~I were more than
twice as large; in Marsden's (1989) follow-up paper the 1970 comet was
classified as a member of a new subgroup, called IIa.

Although the comet's orbit first looked~like~an~oddball, the further
developments showed that it signaled~the~intricate structure of the Kreutz
system, much~more~complex than the two subgroups initially suggested.~Remaining
faithful to the constraint on perihelion~fragmentation, Marsden (1989)
struggled to account for the orbital~relations among the
sungrazers~by~means~of~the~indirect~planetary perturbations
dominated~by~Jupiter.~Ominously, unrealistic orbital periods $<$400~years
had to~be~used;~this compression of the
time scale was necessitated in part by the effort aimed at fitting
Aristotle's comet into the overall hypothesis.  Marsden (1989) admitted
that~his ``{\it scenario \ldots \,is obviously extremely speculative,
and the innumerable free parameters \ldots \,make it inappropriate to
attempt a more rigorous computation.\/}  From the diction of this and
the final paragraph it seems that Brian was a little disappointed with
the results, complaining about being ``{\it \ldots \,painfully aware
that the sequence outlined here requires that six of the eight
intermediaries that apparently passed perihelion unrecorded \ldots
\,ought to have been truly spectacular objects\/}."{\vspace{0.1cm}}

Data from SOHO were a game changer.  By the end of 1999 nearly 100
dwarf Kreutz sungrazers were discovered in the fields of view of
the C2 and C3 coronagraphs, serving as a modest database for early
investigations.  Even though the quality of astrometric observations
from C2 images was not very good and from C3 images was downright
pitiful, limited information could be obtained from the sets of
crude orbital elements that Marsden was publishing first in the
Minor Planet Electronic Circulars (MPECs) and later (sometimes
slightly edited) in the Minor Planet Circulars (MPCs); for many of
the early dwarf sungrazers the orbits also appeared independently
in the International Astronomical Union Circulars (IAUCs).  The
method of astrometric-position reduction was improved by introducing
a new system in early 1999:\ the orbits of all dwarf sungrazers were
recalculated using positions derived by the method of plate constants,
starting with MPC~33650 (1999 Feb 2).

At the very beginning of the 21st century the research of the Kreutz
system was rapidly picking up steam.  In 2001--2002 four major papers
appeared almost simultaneously.  Hasegawa \& Nakano (2001) published
a list of 24~historical comets between 5~BC and AD~1702 that were
deemed possible members of the Kreutz system.  The authors used
mostly Chinese, Japanese, and/or Korean (but for seven objects also
European) sources to derive probable perihelion times.  Interestingly,
Hasegawa \& Nakano's list did not include comet C/1689~X1, mentioned
in Sections~1 and 2.

England's (2002) search for historical records produced 62~candidates,
each of which was tested by applying nine criteria to rank the chance of
its being a member of the Kreutz system.  The three comets of Marsden's
(1967) 17th century cluster achieved the highest rank~9 (including
C/1689~X1), while Aristotle's comet and comet X/1106~C1 reached rank~7.
Unexpectedly, two comets of the 5th century were classified as high as
rank~5 and became a subject of imminent interest.

In the meantime, Strom (2002) examined historical Chinese records of the
Sun's activity and identified 14~instances of the appearance of a star in
the Sun's close proximity between AD~15 and 1865, mostly in the months
of July--September.  Strom speculated that most, if not all, of these
objects were images of daytime comets near the Sun and at least some of
them could be members of the Kreutz system.

I wrote a paper (Sekanina 2002), which --- in addition to other issues ---
showed (i)~the abyss between the most spectacular sungrazers and the dwarf
members of the Kreutz system in terms of intrinsic brightness{\vspace{-0.03cm}}
(a factor of 10$^8$) and (ii)~the dramatic influence that the orbital
location of a breakup had on the motions of the fragments, especially
the contrast between near-perihelion and near-aphelion events.  Most
importantly, a {\it brand new hypothesis\/} of the history and orbital
evolution of the Kreutz system was proposed, which anticipated an early
breakup of the {\it progenitor\/} into two {\it superfragments\/} at
large heliocen\-tric distance, followed by their {\it cascading
fragmentation\/}.  This was a conceptual birth of the two-superfragment
model and the definite end of the notion that sungrazers fragment {\it
exclusively\/} at perihelion.

\begin{figure*} 
\vspace{0.22cm}
\hspace{-0.15cm}
\centerline{
\scalebox{0.795}{
\includegraphics{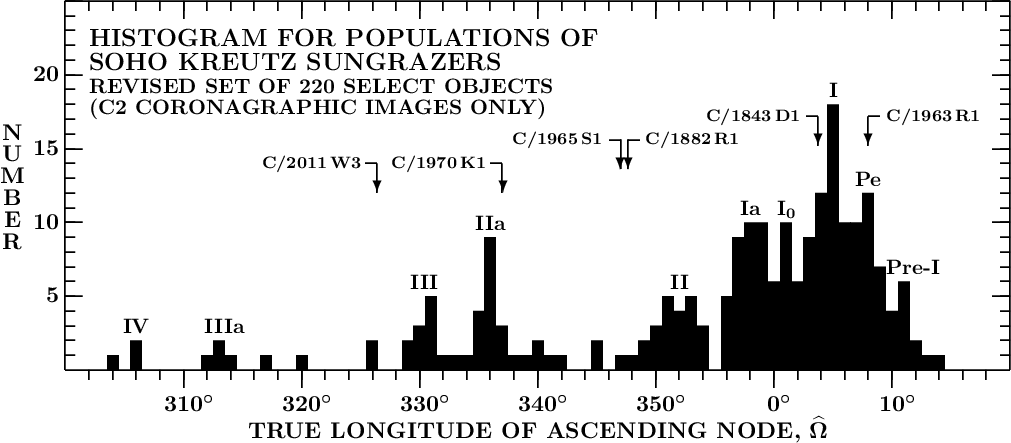}}}
\vspace{-0.1cm}
\caption{Histogram of the true longitudes of the ascending node for a
 set of 220 select SOHO Kreutz sungrazers, whose orbits were derived
 exclusively from the C2 coronagraphic images.  The 10 populations are
 marked.  Populations~Pe and I$_0$ are considered side branches of
 Population~I.  Also depicted are the nodal longitudes of six naked-eye
 sungrazers associated with Populations~I, Pe, II, IIa, and III.  They
 coincide within at most a few degrees with the peaks of the SOHO
 sungrazers. (Adapted from Sekanina 2022c.){\vspace{0.5cm}}}
\end{figure*}

When implemented (Sekanina \& Chodas 2004), the two-superfragment model
was an ad hoc construct that left out dwarf sungrazers and incorporated
a number of assumptions and parametric allowances with major implications,
some of which became over time dubious or downright incorrect [e.g.,
Superfragment~II, equated with X/1106~C1, turned out to be the direct
parent to C/1882~R1 and C/1965~S1 as well as an indirect parent to
C/1963~R1 (of Marsden's Subgroup~I) and C/1970~K1 (of his Subgroup~IIa);
peak separation{\vspace{-0.035cm}} velocities of fragments exceeding
10~m~s$^{-1}$ were allowed; etc.].  Yet, the model was a success at
least in the sense that it demonstrated for the first time the general
feasibility of orbital transfer from one sungrazer to another.  Besides,
the model was able to get all known major fragments into their orbits
in less than two millennia, i.e., slightly more than two revolutions
about the Sun, reckoned from the time of the progenitor's initial
breakup. 

The follow-up paper (Sekanina \& Chodas 2007) dealt with controversial
aspects of cascading fragmentation of the Kreutz system.  Comet X/1106~C1
was now assumed to be a member of Subgroup~I and previously at perihelion
in the 5th century AD as the comet~of~423~or~467, which were ranked rather
high as possible Kreutz sungrazers by England (2002).  Unfortunately, the
flip side of these scenarios was that Aristotle's comet was ruled out as
progenitor (Section 3.2) and the next return of the 1106 comet was
predicted to occur long before the arrival of comet C/1843~D1.  It
appears that neither of the two 5th-century comets was a member of
the Kreutz system, even though they could be its siblings in the sense
that they had separated from the common progenitor at perihelion
{\it before\/} the birth of the Kreutz system.  Nor did the two 5th-century
comets help understand the existence of the eight clusters of possible Kreutz
sungrazers between 1564 and 1970, used in the paper to predict the arrival
of another cluster in the 21st century.

The 2007 paper had a complicated history.  Although its first version was
submitted for publication already in late 2004, it subsequently was being
heavily edited over a period of at least two years.
It so happened that in 2004 Marsden was asked by the editors of
the {\it Annual Review of Astronomy and Astrophysics\/} to review the
history of, and the recent progress in, the investigation of sungrazing
comets in a paper scheduled to appear in the September 2005 issue.  Brian
was determined to update his review as much as possible and include the
conclusions of our research that effectively was still in progress.  I
tried to accommodate his requests for more information as much as I could,
but I may not have been quite successful.  In any case, Brian's paper
suggests that he may have misunderstood some of our points (Marsden
2005).  In retrospect, it is unfortunate that the need to comply with
the review's deadline might have gotten Brian under pressure and precluded
better coordination of his work with ours.
 
Marsden's personal megaproject of the SOHO comets' orbit determination,
which he was dilligently pursuing in addition to all his other commitments,
continued flawlessly.  By the time of his untimely death in 2010, the
set of the Kreutz sungrazers with a known orbit exceeded the impressive
number of 1500.  The orbits were approximate because of the already noted
poor astrometry, and in a plot of the orbit inclination,
$i$, against the longitude of the ascending node, $\Omega$, the SOHO
Kreutz sungrazers displayed a strongly anomalous distribution (e.g.,
Figure~1 in Sekanina \& Kracht 2015).  They lined up along a curve that
made an angle~of~\mbox{15--16}$^\circ$ with the curve, along which the
bright Kreutz comets were distributed and which satisfied the condition
of fixed apsidal line:
\begin{equation}
\tan i = \tan B_\pi \csc(L_\pi \!-\! \Omega),
\end{equation}
where $L_\pi$ and $B_\pi$ are, respectively, the constant ecliptic
longitude and latitude of perihelion.

The anomaly was caused by the normal component~of the {\it sublimation-driven
nongravitational acceleration\/},~$A_3$ (Sekanina \& Kracht 2015, Kalinicheva
2017).  In extreme cases $A_3$ was nearly comparable to the solar
gravitational acceleration.  Because of this effect, $B_\pi$ computed
from Marsden's gravitational orbit was not constant, but depended linearly
on $\Omega$, with \mbox{$dB_\pi/d\Omega = 0.28$}.  For a select set of SOHO
Kreutz comets, whose orbits were computed {\it exclusively\/} from C2 images
(yielding astrometry of better quality), the scatter dropped and the plots of
$B_\pi(\Omega)$ suddenly revealed what turned out to be the actual structure
of the Kreutz system, illustrated~\mbox{in Figure 1.  It is} remarkable that
Marsden's approximate, gravitational orbits, which he himself deemed of
little value, did actually contain important information when based on the
C2 data.  They showed that the longitude of the ascending node represented
a discriminating criterion for sorting out the divisions in the sungrazer
system, as the SOHO comets were distributed unexpectedly tightly along
discrete parallel lines, separated by gaps.  The {\it nominal\/} values
of $\Omega(B_\pi)$ from{\vspace{-0.09cm}} Marsden's orbits along each line
gave the same {\it true\/} value, $\widehat{\Omega}$, for a {\it standard\/}
latitude, $\widehat{B}_\pi$.  In the direction{\vspace{-0.01cm}} perpendicular
to the lines \mbox{$B_\pi = f(\Omega)$} the{\vspace{-0.085cm}} distribution of
$\widehat{\Omega}$ offers the histogram in Figure~1.

This histogram demonstrates that the Kreutz system consists of at least
eight populations with two branches, or of ten populations (or subgroups,
to use Marsden's term), if the branches are counted as independent.  In
the first approximation, the populations --- the side branches excepting
--- are separated{\vspace{-0.01cm}} from each other by gaps of about
equal magnitude, nearly 10$^\circ$.  In retrospect, comet
White-Ortiz-Bolelli was an early warning in 1970 that the division
into two subgroups underestimated the complexity of the Kreutz system.
Forty-one years later, comet Lovejoy (C/2011~W3) further expanded the
orbital diversity among the bright sungrazers.  In terms of the longitude
of the ascending node, the Kreutz system appears from Figure~1 to cover
$\sim$70$^\circ$, which is about 3.5~times wider than the separation of
Marsden's Subgroups~I and II.  On the other hand, these subgroups or
populations do to this day remain the two most prominent ones and the
memberships of the other populations appear to be dwindling{\vspace{-0.07cm}}
in the directions of either end of the total range of $\widehat{\Omega}$; the
rate of the drop is clearly much steeper toward the maximum longitude
at approximately 15$^\circ$.

To comply with the seemingly contradictory properties of the distribution
of the populations, a model should preferably postulate a breakup into two
major fragments, either of them serving as a parent in the continuing
process of gradual disaggregation.  This sounds familiar --- is it not
a two-superfragment model all over again?  One of the particularly
improbable properties of that model was that the main fissure halving the
object went right through the center of the body, along the maximum
dimension.  The way to avoid such a rather embarrassing property is obvious
--- the progenitor ought to be a contact binary, which indeed is most
likely to split in the middle into its two lobes.  This figure has a very
low mass-to-maximum dimension ratio and it has also been gaining support
by observational evidence.  The intrinsic cohesion or propensity for
fragmentation of each of the two objects that became the lobes of the
contact binary determined the distributions of the populations,
respectively, to the left of Population~II and to the right of
Population~I in Figure~1.  Furthermore, the history of the neck
connecting the lobes is linked to the issue of Population~Ia, a component
of the Kreutz system between Populations~I and II that is a little
elusive.

These comments demonstrate that a contact-binary model differs
substantially from the two-superfragment model and is clearly superior
in accommodating the orbital properties of the Kreutz system.  Another
strong point is that, as presented, the contact-binary model is a
pyramidal --- rather than an ad hoc --- construct, compatible with the
observed attributes of the distribution of sungrazer populations in
Figure~1.{\vspace{-0.25cm}}

\begin{figure}[t] 
\vspace{0.15cm}
\hspace{-0.15cm}
\centerline{
\scalebox{0.76}{
\includegraphics{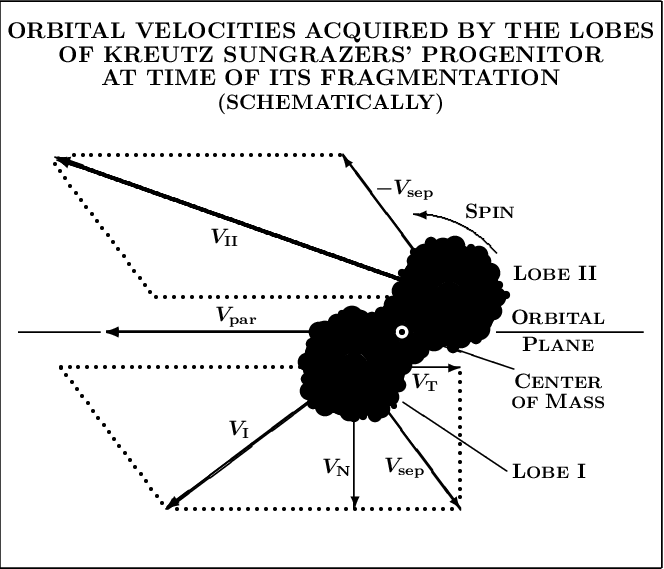}}}
\vspace{0cm}
\caption{Schematic representation of the progenitor (parent) of the
 Kreutz system, modeled as a rotating contact binary, consisting of
 Lobe~I, Lobe~II, and the connecting neck.  The view, at the time of
 aphelion passage, when the lobes separated from each other, is from
 the direction of the Sun and the position of the orbital plane is
 defined by the pre-breakup orbital velocity vector \boldmath $V_{\bf
 par}$.  The dot in the middle of the neck is the position of the
 center of mass of the progenitor, coinciding with the projected
 rotation axis, which is assumed to point at the Sun.  As they
 separated, the lobes began to move in different orbits.  Lobe~I was
 released to the lower right, along the separation velocity vector
 \boldmath $V_{\bf sep}$, while Lobe~II was released to the upper
 left, along the vector \boldmath $-V_{\bf sep}$.  The separation
 velocity consists of the transverse, \boldmath $V_{\bf T}$, and
 normal, \boldmath $V_{\bf N}$, components; the radial component is
 assumed to be \boldmath $V_{\bf R}$~=~0.  Relative to the Sun, the
 orbit into which Lobe~I got inserted is defined by the velocity
 vector \boldmath $V_{\bf I}$, the orbit of Lobe~II by the velocity
 vector \boldmath $V_{\bf II}$.  The contact-binary model is based
 on the premise that the Great March Comet of 1843 was the largest
 surviving mass of Lobe~I and the Great September Comet of 1882 the
 largest surviving mass of Lobe~II.  For the sake of clarity, the
 velocity vectors have not been drawn to scale. (Reproduced from
 Sekanina \& Kracht 2022.){\vspace{0.45cm}}}
\end{figure}

The essentially spontaneous splitting of the contact-binary progenitor
into the two lobes --- with the neck separating either as a third body
simultaneously, or from one of the lobes later --- is of course the
moment of birth of the Kreutz system.  A schematic representation of
this event and its aftermath is displayed as seen from the Sun's
direction in Figure~2, with a detailed description in the figure's
caption.  A realistic model of the Kreutz system is obtained if
the lobes separated at a velocity of about 3~m~s$^{-1}$ near
aphelion, at a heliocentric distance of approximately 160~AU.
Presumably, the separation velocity was rotational in nature.  A maximum
dimension $a$ of the progenitor would imply a rotation period of
\mbox{$P_{\rm rot} = 0.145 \, a$ hr}, e.g., 11.6~hr for \mbox{$a = 80$ km}.
To get a fairly compact swarm of fragments at the time of perihelion
in AD~363 (Section~3.3), the rotation axis of the contact binary must
have pointed at the time the lobes separated essentially at the Sun.
The radial component of the separation velocity was then very close
to zero.  A spread over a period of \mbox{4--5}~days in the swarm
of fragments upon its arrival in AD~363 would be an effect of the
velocity's transverse component.

\begin{figure*}[t] 
\vspace{0.2cm}
\hspace{-0.17cm}
\centerline{
\scalebox{0.875}{
\includegraphics{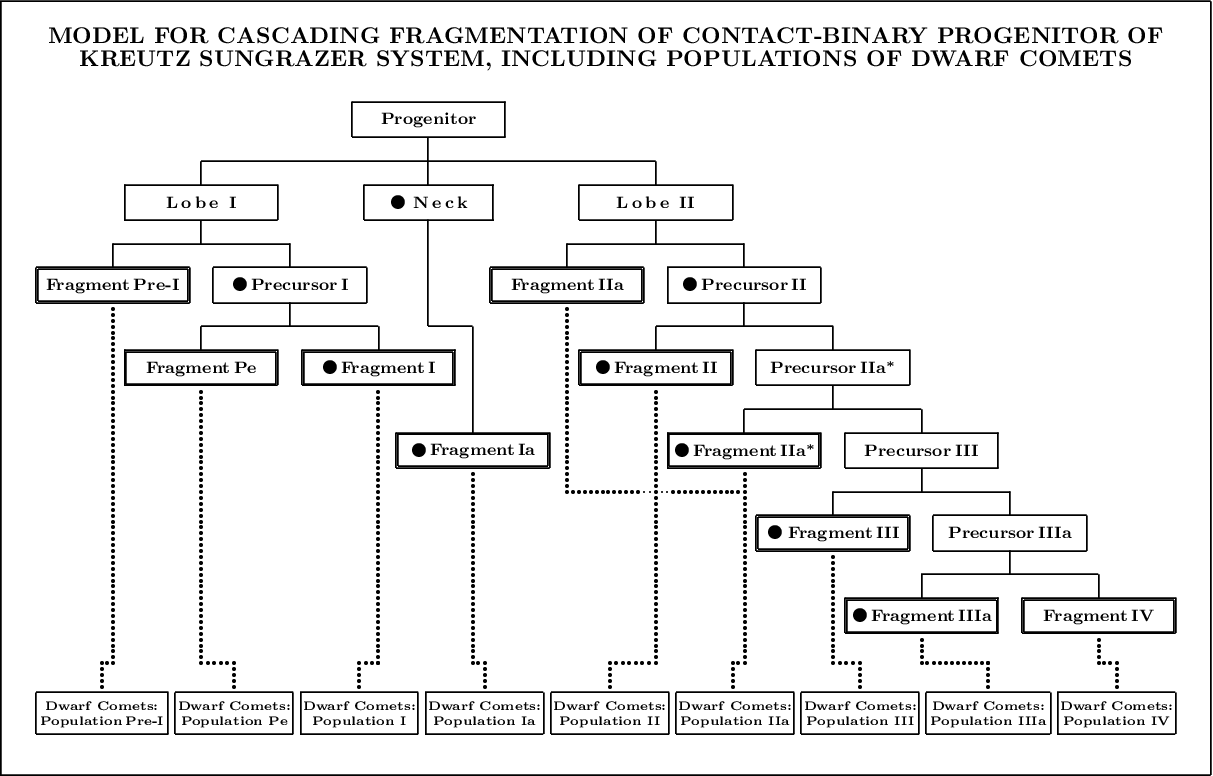}}}
\vspace{-0.05cm}
\caption{Pedigree chart for cascading fragmentation of the model
 contact-binary progenitor of the Kreutz system.  A bullet depicts
 a massive sungrazer that moves essentially in the same orbit as
 its immediate parent.  A sungrazer in a heavily framed box
 successfully reached the first perihelion in AD~363.  The
 dotted lines represent fragment generations between the first
 generation and the last generation --- the stream of dwarf
 sungrazers (such as the SOHO objects) --- at the foot of the figure.
 Population~I$_0$ from Figure~1 is included in Population~I,
 while the separate birth conditions of Population~IIa and its
 branch IIa$^\ast$ are marked. (Adapted from Sekanina
 2022a.){\vspace{0.5cm}}}
\end{figure*}

The solution depends very weakly on the comet's position in the orbit
at the times of the sequence of fragmentation events depicted in the
pedigree~\mbox{chart in Figure~3}.  If the events occurred a century
or so before or after aphelion, the required separation velocities
might be at most 10 or so percent higher.  Similarly, the size of
the gaps that separate two consecutive events could vary
widely, with hardly any effect on the appearance of the swarm of
the daylight comets recorded in AD~363, if the other solution
parameters are properly adjusted.  The only critical parameter is
the radial component of the separation velocity.  If not consistently
near zero, the ``swarm'' would stretch over many weeks or even
months in contradiction to the existing narrative.

\mbox{The$\:$pyramidal$\:$architecture$\:$of the contact-binary$\:$mod-} el
could hardly be brought out more distinctly~than~it~is in Figure~3.  The
first-generation Fragments~I, II,~Pre-I, Pe, Ia, IIa, III, IIIa, and IV
are linked directly to the respective streams of dwarf sungrazers at the
foot of the figure.  However, Population~IIa$^\ast$, a branch containing
fragments whose perihelion distances were much smaller than comet
White-Ortiz-Bolelli's 1.9$\:R_\odot$, shares~the~same stream of dwarf
sungrazers with Population~IIa, while the first-generation
Fragment~IIa$^\ast$ must have differed from the first-generation
Fragment~IIa in both birth and history, and this is reflected in
Figure~3.  On the other hand, this kind of relationship does not
apply to Fragment~I$_0$, which in terms of birth parameters looks
nearly indistinguishable from Fragment~I of Population~I.

\subsection{Aristotle's Comet of 372~BC: The Progenitor}
Even though this extraordinarily brilliant comet mentioned by Aristotle
and other Greek philosophers of the time has often been a very popular
candidate for the earliest member of the Kreutz system, incomplete
information has introduced much uncertainty into the data needed for
successful modeling.

A standard source of information on the comet has over the past two
centuries or so been Pingr\'e's (1783) {\it Com\'etographie\/}, which
unfortunately begins the text with the words by Diodorus Siculus.  Written
about three centuries after the comet's appearance, his description
blatantly contradicts the narrative by Aristotle, who may have witnessed
the comet as a youngster of age 11 or 12.  Compared to Aristotle's text,
Diodorus appears to have deliberately changed the name of the incumbent
Athenian chief magistrate,\footnote{Besides the Olympiad count, the
magistrate's (the~eponymous archon's) name was the basis for the calendar
in ancient~Greece.{\vspace{0.12cm}}} thereby moving the time of the
comet's sighting forward by one year.

\begin{table*}[t] 
\vspace{0.15cm}
\hspace{-0.22cm}
\centerline{
\scalebox{1}{
\includegraphics{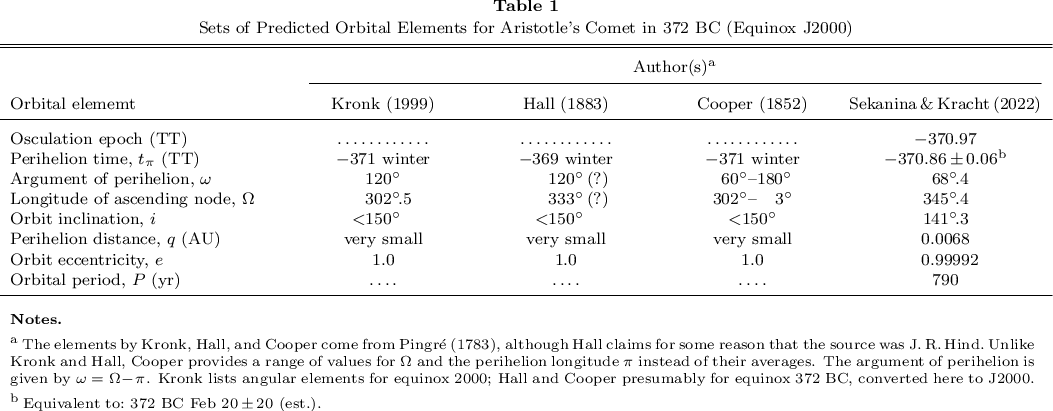}}}
\vspace{0.5cm}
\end{table*}

I address these and related issues in greater detail in a paper that also
shows that the observational and orbital constraints allow one to determine
the perihelion time of Aristotle's comet to better than $\pm$1~month
(Sekanina 2022b).  The constraints on the comet's other orbital elements
have been derived essentially from the narrative by Pingr\'e (1783) in
his {\it Com\'etographie\/}.  The presentations by three investigators,
starting from the same information, are compared in Table~1 with the
results of the orbital integration performed in the context of the
contact-binary model (Sekanina \& Kracht 2022).  The~numbers listed in
Cooper's (1852) catalogue demonstrate that the bounds of uncertainty of
the orbital elements estimated from Aristotle's text are consistent
with the orbital solution derived from the model (the last two columns of
the table), an amazing feat!  Cooper's catalogue includes extensive notes,
which in the case of Aristotle's comet contain some very helpful comments
of his own as well as a translation of segments of Pingr\'e's narrative.

Experience gained by the extensive experimentation with the orbital
integration of suspected historical sungrazers has led to an important
conclusion that if any 5th century object (such as the comet of 423 or the
comet of 467) turned out to be a member of the Kreutz system, Aristotle's
comet had to be ruled out as its progenitor (Sekanina \& Chodas 2007),
replaced instead in some cases with the unremarkable comet of 214 BC.
Conversely, if~\mbox{Aristotle's} comet was the progenitor, no 5th century
object could be a member of the Kreutz~\mbox{system}, although, curiously, it
could be a fragment of Aristotle's comet, moving in an orbit indistinguishable
from the orbit of a Kreutz sungrazer (Sekanina 2023).  This circumstance
was among the points behind recently abandoning the emphasis on the 5th
century sungrazers, introducing instead a contact-binary model (Sekanina
2021), which accommodates Aristotle's comet as progenitor.  Yet, sungrazing
nature of one or more comets in the 5th century (Mart\'{\i}nez et al.\
2022) need not be ruled out.

\subsection{Spectacular Daylight Comets of AD 363:\\Fragments of
 the First Generation}
No more than six words were recorded for posterity by the Greek historian
Ammianus Marcellinus (Rolfe 1940) in his description of a possibly
unprecedented celestial event in late AD 363:\ {\it In broad daylight
comets were seen\/}.  Yet, the ramifications of this statement for the
understanding of the Kreutz sungrazing system are material for a number
of reasons.

First of all, there is the issue of timing, linked to the periodicity of
the Kreutz sungrazers:\ without the 363 event, the gap between Aristotle's
comet and what looked like another very massive sungrazer in the 12th
century was 1477~years, while the subsequent return of a spectacular comet
of the next generation, was another 737~years later, its orbital period
having amounted to just about 737~years as well.  Thus, the first gap was
almost exactly twice as long as the second gap and it also equaled twice
the orbital period, strongly suggesting a missed return.  The Ammianus
event took place 734~years after Aristotle's comet, fitting in just about
perfectly.

Second, the plural --- {\it comets\/} rather than {\it a comet\/} ---
adds to the event's importance in two ways.  One, since the separation
of the two lobes of the contact binary (and their continuing fragmentation)
is modeled to have taken place only a fraction of the orbital period
before 363, the comet is expected to have returned to perihelion as a
swarm of fragments, as there was no time for their dispersal along the
orbit.  And two, since the appearance of several daylight comets is
per se much less statistically likely than the appearance of a single
daylight comet, the perfect timing is by that margin indicating virtually
no chance of some sort of a coincidence.

Third, the appearance in broad daylight implies extraordinarily
brilliant objects and their likely small distance from the Sun both
in the sky and in space.  One immediately suspects sungrazers in
general and the Kreutz sungrazers in particular.  This inference is
supported by the enthralling way that led me to recognize the significance
of this remarkable event~in~the~first place.  It was Seargent's (2009)
remark on~\mbox{Ammianus'} report that attracted my attention.~He observed
that~at~the~time~of the year (say, November--December) Kreutz~\mbox{sungrazers}
``{\it would have had a strong southerly declination and might have
been seen \ldots \,only in the daytime close~to~perihe\-lion.}''
This example demonstrates that the Kreutz sungrazers spontaneously cross
one's mind upon reading the words by Ammianus.

Fourth, even though the lack of detail in Ammianus' narrative is rather
regrettable, there may be reasons why the information on the number of
comets and the extent of their appearance over the sky and in time is
missing.  Intuitively, Ammianus' words seem to imply that more than one comet
was seen {\it at the same time\/}.  If so, there is to be a relationship
between the number of fragments and the extent of their distribution in
space and time.  A crude assessment of these correlations is obtained
from the following exercise.

Let there be $n$ fragments distributed up to a heliocentric distance $r$
along a perihelion arc of a Kreutz orbit approximated by a parabola.
The time from perihelion, $t \!-\! t_\pi$ (in days), varies with the
heliocentric distance (in AU) according to the well-known formula:
\begin{equation}
t \!-\! t_\pi = \mp \:\!27.4 \:\! (r \!+\! 2q) \sqrt{r \!-\! q},
\end{equation}
where the minus sign applies before perihelion, the plus sign after
perihelion, and $q$ is the perihelion distance in AU.  Approximating a
Kreutz orbit by \mbox{$q \rightarrow 0$}, the times of the first and
last fragments relative to the perihelion time $t_\pi$ are, respectively,
\mbox{$t_1 \!-\!t_\pi$} and \mbox{$t_n \!-\! t_\pi$}; the time interval
\mbox{$t_n \!-\!t_1$} (in days) equals
\begin{equation}
\Delta t = \lim_{q \rightarrow 0} (t_n \!-\! t_1) = 54.8 \, r^\frac{3}{2} \!.
\end{equation}
I now vary the solar elongation $\epsilon$ of the first and last~fragments
that corresponds to their distance $r$ and compute the implied time interval
$\Delta t$ equal to the expected duration of the event.  The relation
between $\epsilon$ and $r$ involves the projection of the comet's path as
seen from the Earth.  I approximate the direction of motion of the comet
by the direction of its line of apsides, given by an ecliptic longitude
$L_\pi$ and latitude $B_\pi$.  If the Earth's longitude measured from the
Sun at the time of observation was $L_\oplus$, the angle $\theta$ that
the apsidal line made with the Earth's direction is given by
\begin{equation}
\cos \theta = \cos B_\pi \cos(L_\pi \!-\! L_\oplus)
\end{equation}
and the heliocentric distance is related to the solar elongation by
\begin{equation}
r = \frac{r_\oplus \sin \epsilon}{\sin(\epsilon + \theta)},
\end{equation}
where \mbox{$r_\oplus \simeq 1$ AU} is the distance Sun-Earth at observation.
The event's expected duration is therefore approximately equal to (in days)
\begin{equation}
\Delta t = 54.8 \! \left[\frac{\sin \epsilon}{\sin(\epsilon + \theta)}
 \right]^{\!\frac{3}{2}} \!\!.
\end{equation}

\begin{table}[t] 
\vspace{0.11cm}
\hspace{-0.2cm}
\centerline{
\scalebox{0.99}{
\includegraphics{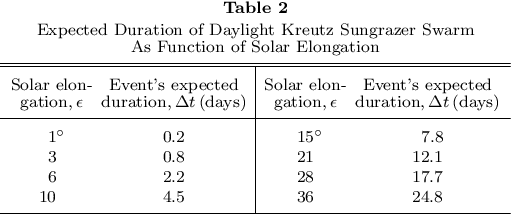}}}
\vspace{0.6cm}
\end{table}

For the estimated observation time of 363 November~15 I find
\mbox{$L_\oplus = 54^\circ\!.1$} and from Sekanina \& Kracht
(2022) the average values for Lobe~I and Lobe~II: \mbox{$L_\pi =
282^\circ\!.7$} and \mbox{$B_\pi = +35^\circ\!.4$} (all equinox
J2000).  From Equation~(4) it follows that \mbox{$\theta = 57^\circ\!.4$}
and Equation~(6) then provides a relation between the solar
elongation and the event's expected duration which is presented
in Table~2.

The range of solar elongations has intentionally been stretched in
either direction to make sure that a ballpark of plausible values is
covered.  In modern times, rather successful attempts have been made
to detect a sungrazer, such as Ikeya-Seki in 1965, only a few degrees
from the Sun (e.g., Millon 1966; Figure~1 in Sekanina 2022a).  However,
a daylight observation of this kind is hard to expect in ancient times
from people who saw the comets because their appearance was overwhelming
and impossible to overlook, not because of observation-facilitating
arrangements made in anticipation of possible detection (such as
looking for a location where the Sun's rays are blocked by a building,
etc.).  For these reasons and also because of an inevitable crowding
of comets over a short orbital arc, I do not consider the first two
entries in Table~2 as feasible.  On the other hand, arguments against
solar elongations exceeding $\sim$15$^\circ$ are even stronger, as the
fading of comets with increasing apparent distance from the Sun is
known to be steeper than the rate of naked-eye detection.  The
conclusion is that the most likely area of the sky for seeing the
daylight comets (with extremely small perihelion distances) was
between 5$^\circ$ and $<$15$^\circ$ from the Sun and a probable duration
of the sightings was 2 to 7~days.  I modeled a case with an extent of
4.6~days (Sekanina 2022a).{\vspace{-0.2cm}}

\subsection{A Comet of 1041:\ Parent to C/1963~R1 (Pereyra)?}
The original barycentric orbital period of Pereyra's comet was
904\,$\pm$\,17~yr (Marsden et al.\ 1978, Marsden \& Williams 2008),
so that its predicted previous appearance was in AD~1059\,$\pm$\,17.
Hasegawa \& Nakano (2001) listed a possible sungrazer in AD~1041, at
almost exactly 1$\sigma$ from the prediction.  It was seen in China
and Korea, first time on September~1, and remained under observation
for a record period of more than 90~days.

On the other hand, England (2002) listed {\it two\/} Kreutz suspects
in 1041:\ a Korean comet discovered in September, which may also have
been seen in Constantinople, and another Korean comet first detected
between October~28 and November~26, which was also seen from Europe.
The first comet remained visible for more than 20~days, the second for
more than 10 days.  Ho (1962) also lists two separate objects.  However,
neither England nor Ho mentioned that the first comet was also observed
in China, a point made by Hasegawa \& Nakano.

Given the experience with the Great September Comet of 1882, the period
of visibility exceeding 90~days implied by Hasegawa \& Nakano suggests
that the comet may have fragmented at perihelion profusely.  Both the
one-comet and two-comets scenarios are in principle possible and either
makes sense in terms of the relationship with comet Pereyra.{\vspace{-0.2cm}}

\subsection{Great Comet of 1106 (X/1106 C1):\ Presumed\\Representative
 of Population~I and\\Parent to C/1843~D1}
Seen in a large number of countries, both in the West, Middle East, and
Far East, this spectacular comet was reported to have been first sighted,
in broad daylight,~in Belgium on 1106 February 2.  In the Far East
the~comet was first detected, after sunset, a week later,~when~the tail
was apparently as long as 100$^\circ$.  The most reliable records
suggest that the object was under observation until about mid-March.
                     
I find no evidence that the identity of X/1106~C1 with Aristotle's comet
(whether with a period of 1477~years or its aliquot part) was proposed by
anyone {\it before\/} 1843.  Pingr\'e (1783) would have been the obvious
candidate, but he noted no orbital resemblance to Aristotle's comet in
any other comet.  With the arrival of the Great March Comet of 1843 the
situation changed dramatically, and I have seen no better account of
the frantic activity aimed at identifying possible past returns of this
comet than what was presented by Cooper (1852).  I comment on this effort
in some detail in Section~3.11, but one needs to read Cooper's notes on
the 1843 comet to appreciate the immense obsession of many astronomers
in the mid-19th-century with the subject of a single sungrazer that kept
coming back.  Indeed, a number of objects, sometimes including the Great
Comet of 1106 but always the 1843 spectacular sungrazer, were suspected
to be returns of Aristotle's comet, but the proposed orbital periods were
invariably much too short, less than 200~years.

Not until four decades later was an attractive scenario presented by
Hall (1883) on the occasion of arrival of the Great September Comet of
1882.  He began by listing the five comets believed at the time to
have similar orbits:\ Aristotle's comet\footnote{With no effect on his
arguments, Hall incorrectly used 370~BC as the year of appearance of
Aristotle's comet.} and the comets of 1668, 1843, 1880, and 1882.  Hall
realized that in order to obtain for the last comet an orbital period
close to 794~years that Frisby (1883) derived from the observations,
there must have been {\it three\/} revolutions about the Sun since the
time of appearance of Aristotle's comet:\ \mbox{$2252/3 \simeq 751$
years}.  The intervening returns were in the years 310 and 989 for the
comet of 1668; 368 and 1106 for the comet of 1843; 381 and 1131 for
the comet of 1880, and 382 and 1132 for the comet of 1882.  Inspecting
{\it A Handbook of Descriptive and Practical Astronomy\/} (1861) by
G.~F.~Chambers, Hall found no relevant comets recorded in 310, 368,~381,
and 989, some suspects in 1131 and 1132, but the book highlighted
1106 with what Hall called a {\it splendid comet\/}.  He concluded
that the pair of Aristotle's comet and the Great March Comet of 1843
ought to be linked with the comet of 1106, only the previous return
missing.  This story rebuffs a false notion, often presented in
the literature, that the comet of 1106 was first suggested to be a member
of the sungrazer group by Kreutz in his famous trilogy, published
between 1888 and 1901.

In Hall's paper, the issue of the perihelion return preceding that
in 1106 has remained unsolved, as he found no comet appearing in
the year 368.  It is unclear why Hall ignored a key comment (even
though marred by the 1843 vs 1882 issue) that Frisby (1883) made
in the paper that Hall referred to.  In reference to Aristotle's
comet Frisby remarked that the Great September Comet ``may possibly
be its third return, {\it a very brilliant comet having been seen
in full daylight AD~363\/}.''

Besides Seargent (2009), this is the only Kreutz-related reference
to the text by Ammianus Marcellinus that I was able to locate.
Combining Hall's and Frisby's comments, one gets a rudimentary view
of the history of one of the two largest surviving masses of the
Kreutz system, which resembles the history that I independently
reconstructed more fully from wider evidence.  Ignoring the minor problem
with the missing plural in Frisby's remark on the event of 363, the
only comet unaccounted~for in the Hall-Frisby narrative is the subject
of next section, as Hall said very little on the history of the
1882 sungrazer.\,\,\,

\subsection{Chinese\,Comet\,of\,1138:\,Discovering\,the\,Missing\,Link\\to
Population~II}
Given that the largest surviving masses of the progenitor's two lobes in
the contact-binary model are identified with the Great March Comet of
1843 (Section~3.11) and the Great September Comet of 1882 (Section~3.13),
there should be historical records of their previous appearances in
the 12th century.  With the comet of 1106 as the only candidate,
two major questions waiting for the answers were:\ (i)~which of the two
19th-century sungrazers was associated with the 1106 comet and
(ii)~what was the 12th-century comet that the other was linked to.\,\,\,

Solutions were proposed in a recent paper on the orbital history of
Kreutz sungrazers (Sekanina \& Kracht 2022).  The problem of the
relationship between the comet of 1106 and the two 19th-century
spectacular sungrazers was,\,perhaps unexpectedly,\,settled by an
in-depth investigation of the motion of another sungrazer --- the
principal nucleus of comet Ikeya-Seki (C/1965~S1).  It is known from
Marsden's (1967) work that Ikeya-Seki and the 1882 sungrazer were
one comet until the 12th-century perihelion and that Ikeya-Seki
split itself at its 1965 perihelion.  Marsden's orbit of Ikeya-Seki's
primary component was the subject of our 2022 paper.

We first determined that an orbital solution, linking 72~preperihelion
astrometric observations with 45~post-perihelion observations of the
main mass, extending all the way to the last data point on 1966~Jan~14,
85~days after perihelion, provided a result closely confirming the
numbers by Marsden.  We obtained a barycentric orbital period of
849.6\,$\pm$\,1.9~years, implying that the previous perihelion occurred
early in the year 1116.  This is where Marsden stopped.  We continued
and found that by linking the same preperihelion observations with 
42~post-perihelion observations ending on 1965~Dec~24, 64~days after
perihelion, the orbital period shortened, placing the previous perihelion
at the end of the year 1119, 3.8~years {\it later\/}.  Continuing further,
we established that the previous perihelion moved to 1124.4, when the
post-perihelion observations were limited to 36 ending on 1965~Dec~7;
to 1126.4, when limited to 23 ending on Nov~20; to 1133.6, when limited
to 17 ending on Nov~14; and to 1136.6,\,when limited to 12 ending on Nov~6,
16~days after perihelion.\footnote{Preperihelion obsevations alone offered
a useless solution with an indeterminate orbital period.{\vspace{0.05cm}}}
Extrapolation to the time of presumed breakup of the nucleus of Ikeya-Seki
at perihelion, the previous perihelion time came out to be near the year
1140, more than 30~years after the appearance of the 1106 comet!

The obvious conclusion from these computations was that {\it
Ikeya-Seki did not split off from the comet of~1106 and neither did the
1882 sungrazer\/}.  By default,~the 1106 comet must have been a member
of Population~I and the previous appearance of the 1843 sungrazer.  This
inference is in line with Hall's (1883) scenario but contradicts Kreutz's
(1888, 1901) belief that the 1106 comet was identical with the 1882
sungrazer.

The exercise with the orbit of comet Ikeya-Seki answered the question of
the whereabouts of the comet that was its parent and the parent of the
1882 sungrazer only approximately --- it was at perihelion around 1140.
However, a search in Ho's (1962) catalogue for historical comets that
appeared about that time offered almost instantly a plausible
candidate --- No.~403.  Ho wrote that {\it on 1138 Sept~3 a broom
star\,}\footnote{Usual translation of {\it hui\/}; a star with a tail,
like a broom.} {\it was observed in the east.  It went out of sight on
Sept~29.\/}\footnote{As references Ho listed two Chinese sources:\
{\it Sung Shih 29/3a\/} and {\it Hs\"{u} Thung Chien Kang Mu
14/7a\/}.}$^,$\footnote{Under the same number Ho also mentioned an
apparently independent object seen in Japan in the northwest on Aug~27.}

From the brevity of the description it is obvious that the comet was by
no means a spectacle.  Even though~its intrinsic brightness should have
been on a par with the comet of 1106, it was at perihelion at the
``wrong'' time of the year.  Accordingly, the Chinese comet of 1138
is neither among the Kreutz suspects proposed by Hasegawa \& Nakano
(2001), nor on England's (2002) list of early possible Kreutz sungrazers.
However, Hasegawa's (1980) catalogue contains this object under No.~636
and it likewise is included in Pingr\'e's (1783) cometography.

While the reader is referred to Sekanina \& Kracht (2022) for details,
it suffices to mention here that the historical record is consistent
with a perihelion time within days of 1138~Aug~1, if the comet was
seen (before sunrise) from a location close to the Sung Dynasty's
capital.  The comet's discovery then occurred 33~days after perihelion.
It is well-known that a Kreutz sungrazer is extremely hard to observe
from the ground (except in daylight if bright enough) when it passes
perihelion between mid-May and mid-August (e.g., Marsden 1967), as it
then approaches and leaves the Sun from behind.  The parent of the
1882 sungrazer and Ikeya-Seki obviously was this case.  In fact, it
is remarkable that it was detected~at~all.  Accepting for the 1138
comet an absolute magnitude of 2.8 before perihelion and 0.7 after
perihelion, its apparent magnitude was predicted at 2.4 on Sept~3
and 4.1 on Sept~29, 0.8~mag above and just about at an estimated
limiting magnitude, respectively.  At a midpoint of the astronomical
twilight the elevation and solar elongation of the comet's ephemeris
position were 11$^\circ$ and 36$^\circ$~on~the first date and 31$^\circ$
and 60$^\circ$ on the last date.

Besides being fully consistent with the orbital period of comet Ikeya-Seki
corrected for the effect of fragmentation, the timing of the 1138 comet
turns out to be in excellent agreement with Marsden's (1967) prediction
of the previous appearance of the 1882 sungrazer; by integrating Kreutz's
(1891) best orbit for nucleus B (or No.~2 in Kreutz's notation) back
in time, Marsden predicted its perihelion to have occurred in April
1138!  Marsden dismissed the result as uncertain because of the presumed
poor quality of the used visual observations.  Instead, it appears that
not only Kreutz's orbit was excellent, but also that nucleus B was by
far the comet's largest mass.

Another support for the Chinese comet of 1138 as the parent of the
1882 sungrazer and Ikeya-Seki comes from the substantially improved
agreement between the orbits of the two objects integrated back to
1138 (Sekanina \& Kracht 2022) over Marsden's (1967) results based
on his integration back to 1115.

In summary, I feel confident that the Chinese comet of 1138 is the
12th century's ``missing'' major sungrazer of Population~II and the
parent of both the Great September Comet of 1882 and comet Ikeya-Seki.
Although its apparent brightness suffered from extremely unfavorable
observing conditions, moving on the far side of the Sun, it appears
that the comet was on a par with that of 1106 in terms of intrinsic
brightness.  The sequence of known major surviving fragments down to
their third generation is now complete.

\subsection{$\!\!$Comet C/1668 E1:\ A Nuisance and Likely Sibling~of\\the Great
March Comet of 1843}
I already noted in Section 2 that starting in 1702 the comet of 1668
provoked a debate on its presumed identity with other sungrazers.  In
the second half of the 19th century the guessing of sungrazers' identities
became a frenzy (Section 3.11), giving this comet --- again in the midst
of it --- a bad reputation.  Although observed by Cassini (1668) more
than three decades before he initiated the first controversy, the only
input to work on the comet's orbit was provided by a map sent from Goa,
then a Portuguese colony in the Indian subcontinent, to Father {\AE}gidius
Franciscus de Gottignies in Rome.\footnote{Some accounts of the story
mistakenly claim that Gottignies made the observations; in fact they
apparently were made by an unknown Jesuit.{\vspace{0.23cm}}} 

Yet, no orbit was computed until after the appearance of the great
sungrazer of 1843, which looked just like the nearly forgotten
comet of 1668.  Henderson (1843a, 1843b), who got hold of a copy
of the map, took up the task only to find out that the positions
read from the map were so uncertain (with errors of up to more than
1$^\circ$) that the orbital solution was essentially indeterminate.
A formal least-squares fit resulted in a set of elements, presented
in column~2 of Table~3, which did not support the suspicion that
the comet was a sungrazer.  An alternative was to force an orbit he
had for the 1843 sungrazer and solve only for the perihelion time.
Following this line of attack, Henderson obtained a result
displayed in the third column of Table~3.  Over the most favorable
period of sighting, 1668~Mar~9--17, the fit was nearly as good as that
by the optimized solution.  Schumacher (1843), reviewing Henderson's
conclusions, remarked that the sungrazing orbit with a perihelion
time about 24~hours later than adopted would have agreed even
better with the observations.  Both orbits by Henderson were
included by Cooper (1852) and by Galle (1894) in their catalogues.

\begin{table*}[t] 
\vspace{0.17cm}
\hspace{-0.2cm}
\centerline{
\scalebox{1}{
\includegraphics{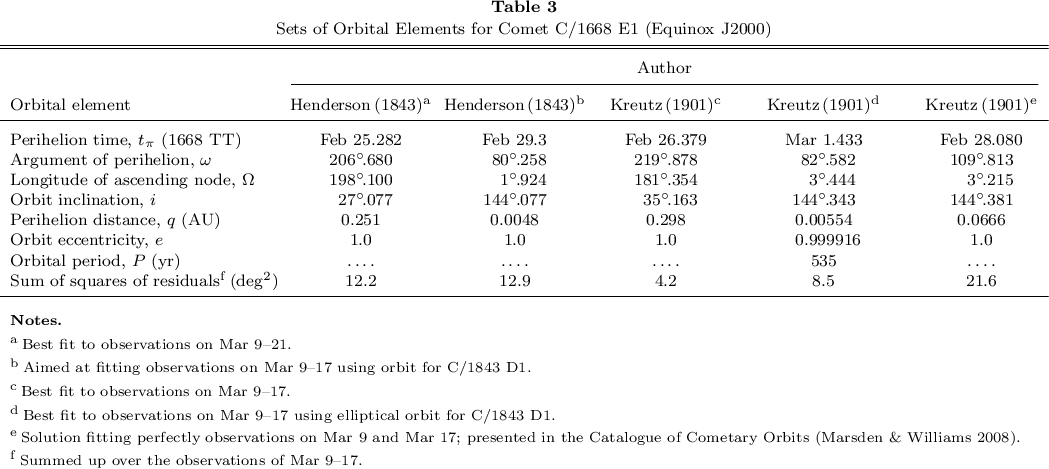}}}
\vspace{0.52cm}
\end{table*}

Nearly sixty years later, Kreutz (1901) essentially repeated
Henderson's work and obtained results that --- as one would expect
--- were not dramatically different.  His best parabolic solution
is shown in column~4 of Table~3, while forcing his elliptic
orbit for C/1843~D1 in column~5.  The last column offers an
orbit that fits exactly the positions on the first and last
days, March~9 and March~17.  This is a very poor solution, the
worst in Table~3, leaving strong systematic trends in the
residuals.  Yet, this was the comet's orbit selected for the
catalogue by Marsden \& Williams (2008).  I admit that the
reasoning behind this choice somehow eludes me.  An important
additional piece of information that Henderson could not
offer was Kreutz's determination that the motion of the 1668
comet was much more consistent with the motion of the Great
March Comet of 1843 than with the motion of the Great September
Comet of 1882.  Quite probably, the comet of 1668 was a major
Kreutz sungrazer of Population~I.{\vspace{0.03cm}}

Brand-new developments reported by Golvers (2025) suggest that
the comet of 1668 was observed by a Jesuit in Macau, as
documented by a fragment of a personal diary, and by another
Jesuit living in Peking, now confirming previously suspected
evidence, based on the contents of a letter (Golvers 2014).
The existence of these documents is in broad agreement with
the statement by Hasegawa \& Nakano (2001) that many records
of the 1668 comet are known in China, as well as Korea and
Japan.  However, I am at present unsure whether this new
information includes any positional data of sufficiently high
quality to contribute to a significant improvement of the
comet's orbit.{\vspace{0.03cm}}

My search in the literature, especially that dated before 1843,
for any suggestions of the possible identity~of the 1668
and 1106 comets turned out to be negative.~This may or may not
have been affected by Cassini's proposed identity of the comets
1668 and 1702 and the implication, mentioned by Schumacher, of
their relationship with Aristotle's comet.  Indeed, the identity of
the comets of 1668 and 1702 implied an orbital period of 34~years,
which multiplied by 60 made it very close to 372~BC.  On the
other hand, these numbers indicate that it was 16.5~revolutions
about the Sun to get from 1668 back to 1106.  One would have to
assume that between 1668 and 1702 the comet made two revolutions
in order to fit in the comet of 1106.  That may have made the
orbital period too short even for Cassini.{\vspace{0.03cm}}

The search was motivated by my strong suspicion~that the 1668
comet was a major fragment of the 1106 comet, potentially its
second largest fragment after the Great March Comet of 1843,
that separated at perihelion.  This issue is of interest
because it may indirectly imply the arrival of a bright
Kreutz sungrazer in the relatively near future, as discussed
in Section~4.

\subsection{Comet C/1689 X1:\ An Unlikely Kreutz Sungrazer}
Reports of sightings at the Cape of Good Hope before perihelion
(on 1689~Nov~24--25) and, possibly, at sea near perihelion
notwithstanding, the comet was seen mainly in the period of
1689~Dec~8--24, with the tail reaching 68$^\circ$ on Dec~14.
Yet, information on its orbital motion was extremely limited
and of poor quality, reflected in highly discordant sets of
elements, apparent from Table~4; the first two sets are in
Cooper's catalogue, all four in Galle's.  Of much concern is
the enormous scatter in the longitude of the ascending node,
given the steep inclination of the orbital plane, as well as
the fact that the four solutions do not even agree on the
direction of the comet's motion.\,\,\,\,

There has been a wide range of opinion about the chance that
this comet was a member of the Kreutz system.  Probably the
most positive attitude was expressed by Kendall (1843), who
--- as noted in Section~2 --- believed that this comet was
identical with the 1843 sungrazer, based on the set of elements
derived by B.~Peirce (Table~4).  This set not only resembled
a sungrazing orbit, but the longitude of the ascending node
essentially coincided with that for Population~II, whose existence
was unknown at the time.  More recently, the comet~was
apparently considered a Kreutz sungrazer by England (2002),
who assigned it the highest possible rank.

On the other hand, Vogel (1852a), the author of another orbit
listed in Table~4, dismissed the idea of the 1689 and 1843 comets
being identical as ``{\it not by any means admissible\/}.''  Also
skeptical about the prospects of the 1689 comet being a member of
the sungrazer system were both Kreutz (1901) and Holetschek (1892),
the author of the orbit included in Marsden \& Williams' (2008)
catalogue.  Trying to fit the comet's observations with the orbit
of the 1882 sungrazer, Holetschek complained that he hit ``{\it
significant obstacles\/},'' while Kreutz remarked that the orbit
of the 1843 sungrazer was even less appropriate for the purpose
than the orbit of the 1882 sungrazer.  Finally, the 1689 comet was
not on Hasegawa \& Nakano's (2001) list of Kreutz suspects.  Early
orbital efforts were extensively reviewed by Plummer (1892).{\vspace{-0.1cm}}

\begin{table*}[t] 
\vspace{0.17cm}
\hspace{-0.2cm}
\centerline{
\scalebox{1}{
\includegraphics{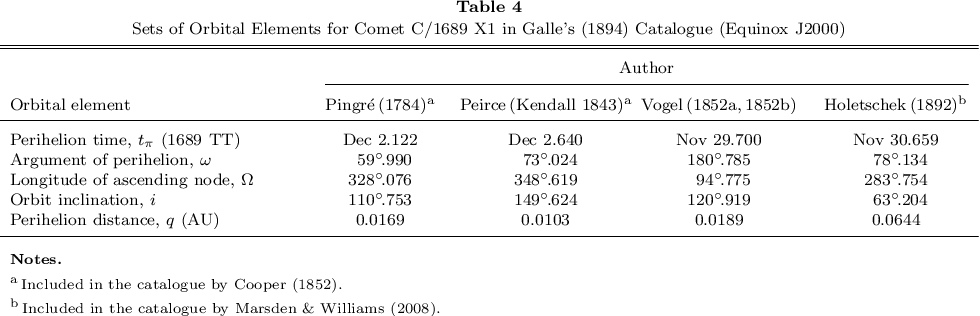}}}
\vspace{0.5cm}
\end{table*}

\subsection{Comet C/1695 U1:\ Another Miss?}
This comet was another example of the obsession with the presumed
previous appearances of the Great March Comet of 1843.  This time
the affected astronomer was von Boguslawski (1845), who in a letter
to Schumacher wrote that the ``{\it previous appearance of the 1843
comet took place in 1695.  Adopting 1695~October~24 as the~time of
perihelion passage, all data on\/} [{\it the comet's\/}] {\it
apparent motion, which Pingr\'e refers to in his Cometographie, are
readily described with the parabolic elements for the 1843 comet
\ldots\/}''  And he continued, ``{\it with an orbital period of 147~years
and 4~months (increased up to 147~years and 9~months, as needed) one
encounters the great comets~of 1548, 1401, 1254, 1106, 367, 219,
and AD~72, and finally the one from 371 BC, whose appearance was
described by Aristotle in the 1st book, chapter~VI of his
Meteorologie.  Of these are the\/} {\small \bf appearances in
1106 and at the time of Aristotle \ldots so unmistakably similar
to that in 1843 that there can be almost no doubt on the identity.}''

Although there is nothing wrong with the text I typed in boldface,\,one
does not need to involve the~1695~comet\footnote{This story
continued (and got worse) after the appearance of comet C/1880~C1
thanks to Weiss (1880); see Section~3.12.} or any of the other comets
between the years 72 and 1548,  as the identity of the comets
of 371~BC, 1106, and 1843 is supported by the fact that the gap
between 371~BC and 1106 is almost exactly twice as long as the gap
between 1106 and 1843.  Interestingly, none of the comets recorded
in the years 72, 219, 367, 1254, 1401, or 1548 is on Hasegawa \&
Nakano's (2001) list of Kreutz suspects.\,\,\,

Kreutz (1901) considered the data on this comet collected in
Pingr\'e's {\it Com\'etographie\/} essentially useless, and von
Boguslawski's letter may have alerted him to initiate an investigation
of the whereabouts of a set of~additional observations.
J.~C.~Burckhardt, a German-French astronomer, found out in 1812 that
numerous observations of the 1695 comet were secured by P.~Patouillet
on Oct~29--Nov~17 (UT) from a ship Le~Floriant when sailing the Arabian
Sea.  Announced by J.\ N.\ Delisle in 1751, the results comprised
a document that was part of his bequest and became eventually
published\footnote{Kreutz (1901) pointed out that when the comet
of 1695 became a subject of increasing interest after the discovery
of the 1843 sungrazer, most astronomers knew about Burckhardt's
elements, but not the observations; this explains why von~Boguslawski
asked Schumacher for help to get them published --- they already were.}
by Burckhardt (1817) together with a set of (very crude) elements.

Kreutz (1901) was apparently the first to exploit these observations
extensively.  Of the 21 data points, eight had to be rejected right
away.  Two of the better orbits that he derived from the remaining
13~observations are listed in Table~5, IIIa --- the first and better
one --- being the set included in Marsden \& Williams' (2008) catalogue.
It is closer to a sungrazing orbit (athough the perihelion distance is
still much too large), but the angular elements are outside the range
of the Kreutz orbits.  Orbit IVa has the longitude of the ascending
node and inclination in the Kreutz range, but it is not even a
sunskirting orbit.  Note that this orbit indicates that the comet
passed through perihelion on exactly the same date as comet Ikeya-Seki
270~years later --- a pure coincidence.

Overall, Kreutz was not inclined to consider the comet a member of
the sungrazer system, because his attempts to fit the observations by
the orbits of the Great March Comet of 1843 or the Great September
Comet of 1882 (solving only for the perihelion time) both left very
large systematic residuals of up to nearly 8$^\circ$.  Comparing
the two he remarked that the unsatisfactory representation of the
observations is even more obvious for the latter than for the former.

England (2002) assigned comet C/1695 U1 the highest possible rank as
a Kreutz sungrazer.  Similarly, Hasegawa \& Nakano (2001) tended to
agree that this was a Kreutz sungrazer, based on the Korean and
Chinese observations.  Koreans recorded it on Nov~3--7, noting that
a few days later it disappeared.  Chinese saw the comet ``after''
Oct~22, that is, potentially before or very near perihelion, and
noted that it was still visible on Nov~14.

\begin{table}[b] 
\vspace{0.4cm}
\hspace{-0.2cm}
\centerline{
\scalebox{1}{
\includegraphics{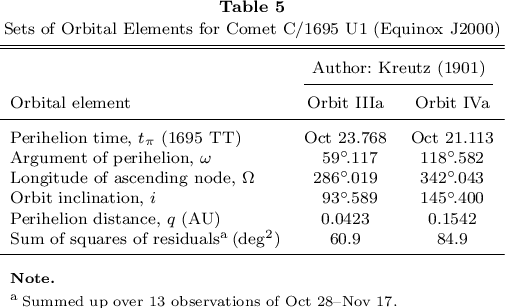}}}
\vspace{0.06cm}
\end{table}

\subsection{Comet X/1702 D1: An Apparent Member\\of Population II}
Unlike in the cases of the comets of 1668, 1689, and 1695, the prefix
for this comet acknowledges that no orbit could be computed.  Indeed,
the comet is not included in Cooper's, Galle's, or Marsden \& Williams'
catalogues.  Yet, Kreutz\,(1901)\,expended much effort~to~learn~as~much
as possible about the motion of the 1702 comet by examining
remarks by three independent observers, referring to the path of
the comet between Feb~27 and Mar~2.  He concluded that these
observations ``{\it can be represented by the orbits of the\/} [{\it
1843 and 1882\/}] {\it sungrazing comets and that, in particular, it
is the orbit of\/} [{\it C/1882~R1\/}] {\it that best satisfies the
tested conditions\/}.''

On the strength of Kreutz's conclusion, supported by England's (2002)
very high ranking of the comet, it is reasonable to argue, in the
context of the contact-binary model, that X/1702~D1 apparently was a fragment
of the Chinese comet of 1138 (Section~3.6) and thus a member of the
Kreutz system's Population~II.{\vspace{-0.15cm}}

\subsection{Remarkable Comet C/1843 D1:~The~Most~Massive Surviving
Fragment of Lobe I}
Numerous daylight observations of this magnificent object at close
proximity of the Sun, made at most merely hours from the time of
perihelion on 1843~Feb~27.91~UT, were a momentous
experience.  The tail reached a maximum length of nearly
50$^\circ$ later in March, when it appeared almost perfectly
straight as a result of the Earth's transit across the comet's
orbital plane on Mar~22.8~TT.\,\,\,

Many 19th-century astronomers were both convinced about, and fascinated
by, the 1843 sungrazer's identity with historical comets.  A nice
account of his own outlook was presented by Cooper (1852), demonstrating
the sheer extent of his involvement in the debate (Section~2).  The
issue at the time was not the likelihood of identity of the 1843 sungrazer
with any previous comet but, rather, with which and how many comets, as
well as what actually was the orbital period.  The undisputed favorite
was the comet of 1668, so that the periods did not exceed 175~years
(e.g., Arago 1843).  Besides the already mentioned proponents, the idea
of two or more objects being the same comet was supported (or accepted
as plausible) by Laugier \& Mauvais (1843a), Nicolai (1843), Capocci
(1843), Valz (1843), etc.  And later on, Laugier \& Mauvais (1843b)
proposed a new, shorter period of 35.1~years,{\vspace{-0.07cm}} close
to $\frac{1}{5}$ of the 175~years, and identified ``similarly looking
comets'' in the years 1843, 1702, 1668, 1528, 1492, 1457, 1106, 1001,
685, 580, 369, 335, 194, 159, and --367.  Aristotle's comet and the
comet of 1106 figured in these scenarios as only two of many candidates.

Cooper's (1852) plan of action was to select 47 historical comets from
Pingr\'e's (1783, 1784) {\it Com\'etographie\/} to serve as candidates
for a previous appearance of the 1843 sungrazer, confronting them with
four potential orbital periods:\ the basic one, given by the appearances
of the 1668 and 1843 comets, and by the three shorter ones, proposed
by von Boguslawski (1845), by Laugier \& Mauvais (1843b), and
Clausen (1843), {\vspace{-0.07cm}}respectively:\ 175, 147$\frac{1}{3}$,
35.1, and 21$\frac{5}{6}$~years.  An elaborate{\vspace{-0.06cm}}
examination of plausible sequences of the candidate objects (whose
likelihood of being sungrazers was weighted) showed that the largest
number of candidates was offered by the period of 35.1~years.

What happened next? Hubbard's (1851, 1852) comprehensive investigation of
the 1843 sungrazer's motion, completed at about the same time as Cooper's
catalogue, indicated a likely orbital period between 530 and 800~years,
profoundly inconsistent with the hypothesis of a single returning object.
And thirty years later, the nucleus of the 1882 sungrazer split into
several fragments at perihelion and the controversial paradigm of
one object's recurring appearances collapsed almost overnight.

\subsection{Comet C/1880 C1:\ The Last Hope for\\the Single-Object
 Believers}
Yet, before the hypothesis of a single returning object was definitively
discredited, the Great Southern Comet of 1880 (C/1880~C1) offered one
last chance to the hopeless cause.  Its orbit turned out to be almost
a copy of the orbit of the Great March Comet of 1843.  All one had to
admit was that the orbital period now was not 35.1~years, but 36.9~years,
not much of a difference.

This indeed was the conclusion that Weiss (1880) came up with, arguing
that the comet of 1843 returned.  He noted that 36.9~years is almost
exactly one quarter of the period proposed by von Boguslawski (1845) and
suggested that the comets of 1695, 1511, 1363, 1179, 1106 were previous
appearances of the 1880 (and 1843) comet.  Weiss was enthusiastically
supported by Meyer (1880), who reported a preliminary orbit for the
new comet derived on the assumption of a 36.9~year period.  To support
this scenario, Meyer cited Plantamour's (1846) result of 21.9~years for
the orbital period of the Great March Comet of 1843.  However, Plantamour
ended his paper with a prediction that the comet would return in the
first days of 1865, a lost cause in 1880.  Subsequently, Meyer (1882)
used a standard method of orbit determination to compute a ``definitive''
set of elements for the 1880 sungrazer; finding that the solution was
indeterminate, he had to bypass the problem to estimate that the orbital
period ranged between 31.5 and 47.7~years, a result that Kreutz (1901)
was unable to confirm.

Then the research appears to have taken a peculiar turn.  Galle (1894)
pointed out that in a special essay on the sungrazing comets of 371~BC,
1668, 1843, and 1880 E.~F.~W.~Klinkerfues affirmed that because of the
considerable proximity to the Sun at perihelion, a sungrazer's motion
could be subjected to a dramatic shortening of the orbital period and
that after 1880 it should further drop to only 17.5~years.\footnote{\mbox{I
have been unable to locate the essay; I did not find a word} \mbox{about
it even in Klinkerfues' obituaries.} \mbox{However, it is easy
to}~re\-construct his prediction of a diminishing
orbital~\mbox{period}.~\mbox{Between} \mbox{1668 and 1843 the
period was 175.0~years,\,equivalent to an\,inverse} \mbox{semimajor axis
of $1/a = 0.0320$\:AU$^{-1}\!$.\,Between 1843 and 1880 the}{\vspace{-0.06cm}}
\mbox{period dropped to 36.9 years. \,Equivalent to $1/a \,=\, 0.0902$
AU$^{-1}\!$,}{\vspace{-0.05cm}} \mbox{this is an increase in $1/a$ by
$0.0902 \!-\! 0.0320 = 0.0582$ AU$^{-1}\!$.\,\,After}{\vspace{-0.06cm}}
\mbox{1880 the $\!1/a$-value should become $0.0902 \:\!\!+\:\!\! 0.0582 =
0.1484\!$ AU$^{-1}\!$,} which indeed is equivalent to a period of 17.5
years.}

In the context of the then popular hypothesis of resisting medium
(e.g., Encke 1836), the issue was addressed by von Oppolzer (1880).
Based on the data available~for comet 2P/Encke and some other short-period
comets, he estimated for a sungrazer an increase in the inverse
semimajor axis to be less than 0.0003~AU$^{-1}$ per revolution
about the Sun.  This is more than two orders of magnitude smaller than
implied by the assumption that the comets of 1668, 1843, and 1880
are one object.

Unlike von Oppolzer, von Rebeur-Paschwitz\,(1884)\,did not wish to
address the issue of whether the hypothesis of resisting medium was at
the root of the potentially diminishing orbital period of the 1880
sungrazer.  Instead, he wanted to find out whether its observations were
reasonably consistent with an orbital period of 17.18~years,\footnote{The
difference between von Rebeur-Paschwitz's and Klinkerfues' values for
the predicted orbital period{\vspace{-0.01cm}} is explained by rounding
off the relevant value of $1/a$ to two{\vspace{-0.06cm}} decimals:\
0.1484~AU$^{-1}$ becomes 0.15~AU$^{-1}$, equivalent to 17.18~years.
One could question retaining the two decimals for the period, but the
discrepancy of 0.32~year made no difference in practice.} which the
post-perihelion motion of the comet should have complied with, if the
hypothesis applied.  His computations showed that in comparison with
a mean residual of $\pm$3$^{\prime\prime}\!$.09 left by Meyer's (1882)
orbital solution with a period between 31.5 and 47.7~years, forcing
a period of 17.18~years led to an orbital solution that left a mean
residual of $\pm$4$^{\prime\prime}\!$.83.  Although larger than Meyer's,
the difference was deemed insignificant enough by von Rebeur-Paschwitz
that he felt the shorter period could not be ruled out.  Kreutz's (1901)
subsequent in-depth analysis of the motion of the 1880 sungrazer led
him to a conclusion that its orbit could not be distinguished from a
parabola.

As pointed out in Section 2, the 1880 comet arrived~at a time when
the new hypothesis of a multitude of objects moving in very similar
orbits was spearheaded as an alternative explanation for the recurring
appearances of sungrazers.  Time was getting ripe for a confrontation
between the two competing scenarios.

\subsection{Stunning Comet C/1882 R1:\ End of Controversy}
The arrival of yet another sungrazer only 31~months after comet C/1880~C1
was a shock, because the time gap between the two was nearly ten times
shorter than the shortest expected period had suggested.  Yet, one guy
(Penrose 1882) managed to graphically determine an early orbit for the
new comet with a period of 480~days!  Although the comet was
discovered and tracked before perihelion, most sets of orbital elements
were determined mainly from the post-perihelion observations.  And while
these results clearly showed that there were fairly large differences
between this and the 1880 comet in both the position of the orbital
plane and the perihelion distance, it was clear that both orbits were
sungrazing.

Hopeful signs of a breakthrough event needed for abandoning the hypothesis
of a single returning object may have appeared as early as 10~hours after
the comet's perihelion passage, but the person was unaware that his
observation was of potentially momentous importance.~I~am talking about
Gill's (1882) remark that he ``{\it was astonished at the brilliancy of
the comet\,}''as it rose in the east a few minutes before the Sun
in the morning of September~18.  Gill continued that ``{\it to my
intense surprise the comet seemed in no way dimmed in brightness\/}''
after sunrise.  To see it in broad daylight ``{\it it was only
necessary to shade the eye from direct sunlight with the hand at
arm's-length\/}.''  Gill was not alone who commented on the extraordinary
brightness of the nucleus.  For example, as part of their spectroscopic
observations with a 24-cm telescope, Thollon \& Gouy (1882) remarked
that the nucleus was very bright and fairly large on September~18.6~UT;
similarly, Wilson (1883) reported it very bright and round in the finder
of the 28-cm equatorial of the Cincinnati Observatory on September~18.8~UT;
etc.  I am aware of no such explicit statements made 24~hours earlier,
hours {\it before\/} perihelion, so that the post-perihelion reports
of a prominent nuclear condensation may suggest a genuine anomaly.

I mention these comments because it is rather common that the breakup
of a comet's nucleus is accompanied by an outburst.  It is conceivable
that the nuclear condensation of the 1882 comet brightened because it
split at perihelion.  However, some cometary outbursts are not associated
with nuclear fragmentation, so that flaring-up is not a fully reliable
breakup signature.

There is another feature that can provide evidence of a recent breakup.
It is well-known that fragments of a split nucleus separate at extremely
low, near-zero velocities, so that it takes a long time before the
distances between the fragments are large enough to detect from the
Earth.  Experience with more recent split comets and improvements in
modeling the motions of fragments show that --- depending on the
spatial resolution of the used telescope --- seldom is their separation
seen earlier than a few weeks after the event.  What is detected earlier
than the individual fragments is a {\it steadily increasing elongation\/}
of the nuclear condensation.

In the case of the 1882 sungrazer, it was communicated
by~de\,Bernardi\`eres\,(1882),\,the leader of the French
Venus-transit$\:$expedition$\:$to$\:$Chile,\,that,\,employing$\:$an$\:$equatorial,
L.~Niesten observed the comet's nucleus to be oval and tilted about 30$^\circ$
to the tail axis on September~21.45~UT, less than four days after perihelion.
Barnard (1883) reported the first detection of an elongated nucleus, in the
direc\-tion of the tail, on September~27.~He~continued~that the
nucleus was slightly more elongated~on~October~1, six or seven times as long
as it was broad on October~4, and still much more elongated and separated
into three unequal parts on October~5.  On the other hand, in the long
postponed publication, Gill (1911) wrote that there was not ``{\it any
appearance of elongation noted during the observations of September~28.
On September~30 \ldots Dr.\,Elkin \ldots with the limited aperture of
the heliometer saw only an elongation of the central or most condensed
part of the nucleus, whilst} [{\it Mr.\,Finlay\/}], {\it with the
6-inch equatorial, figured a decided separation of the central
condensation into two separate points of condensation\/}.''  From
this evidence Gill concluded, somewhat naively, that the ``{\it
disruption of the nucleus and the formation of separate points
of condensation seem therefore to have commenced between
September~28\/$^{\it d}$\/17\/$^{\it h}$ and September~30\/$^{\it
d}$\/17\/$^{\it h}$\/C.M.T.\/}''  Yet it is rather obvious that
just as Elkin did not see the secondary nucleus on the 30th because
of inadequate spatial resolution, neither did Finlay nor Elkin detect
the elongation on the 28th either for the same reason or because of
poor seeing.

More trustworthy were Gill's (1911) remarks on the appearance of the
comet's nucleus before perihelion.  He said that the ``[{\it nuclear\/}]
{\it condensation became a smaller and more sharply defined perfectly
circular disc as the comet approached perihelion\/}'' and that on
September~17, only hours before perihelion, the ``{\it central
condensation or nucleus was certainly single.  Half-an-hour before
its disappearance at the Sun's limb its diameter was~4\,$^{\prime\prime}$
according to Mr.\,Finlay's measures\/}.''\,These descriptions~are
consistent with a breakup at perihelion.

By the first days of October, the presence of two condensations was
detected at several observatories.  The nuclear region eventually
developed into a chain of fragments, which lined up like beads on a
thread of worsted and were enveloped by diffuse material.  Kreutz
(1888) listed nine reports of a double nucleus by October~7.0, 19 by
October~14.0, and 35 by October~21.0~UT.  Three separate condensations
were seen for the first time on October~6 and four on October~13.
Altogether, up to six condensations were detected simultaneously, but
the separation distances among them were not always measured.  The
nuclear train was steadily growing in length and so were the gaps
between the condensations.

The impressive appearance of the fragmenting nucleus of the Great September
Comet of 1882 with the far-reaching ramifications was surely the most
powerful argument to refute the hypothesis of a single returning object:\
if it happened in 1882, it must have happened a number of times before,
and the observed sungrazers were products of such past events of nuclear
fragmentation.  After the arrival of comet Ikeya-Seki in 1965 Marsden
(1967) offered a proof that one such event did undoubtedly take place in
the early 12th century.

Kreutz's (1888, 1891) orbital computations of the 1882 sungrazer
contributed significantly to the speedy acceptance of the new hypothesis.
First of all, by successfully linking the preperihelion astrometric
observations of the single nucleus with the post-perihelion ones of
four nuclear fragments Kreutz showed that the Sun's corona did not
measurably affect the comet's orbital motion and did not change its
period, thereby ruling out any resisting-medium effects.  Second,
by computing high-quality orbits for the four nuclear fragments,
Kreutz showed that their orbital periods were in a range of
600--1000~years, comparable to the orbital period derived by Hubbard
(1851, 1852) for the Great March Comet of 1843,~but~one order of magnitude
longer than those typically assigned to sungrazers by the hypothesis of
a single returning object.  And third, the gaps {\it between\/} the
orbital periods of the four nuclear fragments computed by Kreutz were
in a range of 60 to 100~years, comparable to, and accounting for, the
times that the hypothesis of a single returning object required for
the orbital periods of sungrazers.

Kreutz's (1891) investigation of the comet's orbital motion went
beyond the three points.  One of the second-order issues that he did not
miss was the independent derivation of the nuclear fragments' orbital
elements from (i)~linked pre- and post-breakup astrometric observations
and (ii)~post-breakup observations alone.  The differences between the
orbital periods obtained these two ways allow one to estimate the
orbital period of the single, pre-split nucleus (whose direct computation
failed).  In spite of the fairly large errors, the results in
Table~6 do suggest that~by adding the pre-breakup observations, the
orbital period of Fragment~A increased, whereas the orbital periods of
Fragments~C and D decreased, always shifting in the direction toward
772~years, the orbital period~of~Fragment~B (Point~2 in Kreutz's
notation), which appears to have been moving essentially in the orbit
of the original comet.  This was independent evidence for Fragment~B
as apparently the most massive one among the four.

\begin{table}[t] 
\vspace{0.17cm}
\hspace{-0.21cm}
\centerline{
\scalebox{1}{
\includegraphics{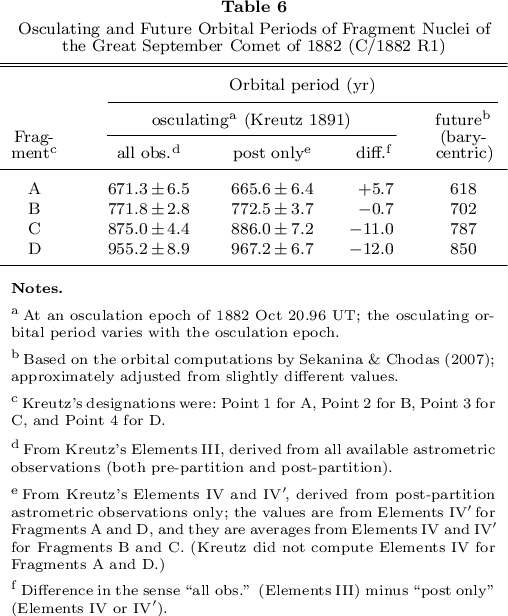}}}
\vspace{0.5cm}
\end{table}

More recently, application of a standard model for the split comets
confirmed that the extensive fragmentation of the nucleus of the 1882
sungrazer took place in close proximity of perihelion (Sekanina \&
Chodas 2007), always $<$\,3~hours after the passage.  Since it took
1.8~hours to the 1882 sungrazer to get from the perihelion point to
the point of +90$^\circ$ in true anomaly, the heliocentric distance
at fragmentation may have exceeded 0.015~AU.  It is unclear whether
the birth times of the individual fragments did actually coincide.

Comparison with the even more extensive fragmentation of comet
Shoemaker-Levy (D/1993~F2) at Jupiter shows common features.
Either object fragmented {\it after\/} the point of closest
approach:\ 3.1\,$\pm$\,0.2~hr, derived from eight (E--W) fragments
of Shoemaker-Levy~(Sekanina et al.\ 1998); and 2.4\,$\pm$\,0.4~hr,
derived from four \mbox{(A--D)} fragments of the 1882 sungrazer
(Sekanina \& Chodas 2007), suggesting the presence of inertia
effects.  Furthermore, unlike in cases of nontidal breakup, the
fragments at both ends of the train (A~and~W in Shoemaker--Levy, and
A and D in the 1882 sungrazer) were fainter and presumably
less massive, whereas the brightest and presumably most
massive fragments (G, K, L, Q in Shoemaker-Levy and B in the 1882
sungrazer) were located near the middle of the train.  The very
existence of cometary fragmentation in close proximity of Jupiter
suggests that~the magnitude of the tidal force, not an effect of the
high-temperature environment, was the decisive factor in these
breakup events.

\begin{table} 
\vspace{0.17cm}
\hspace{-0.21cm}
\centerline{
\scalebox{1}{
\includegraphics{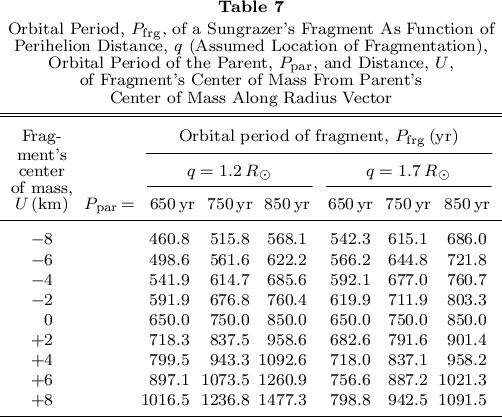}}}
\vspace{0.5cm}
\end{table}

\section{Prospects for Appearance of a Bright Kreutz Sungrazer in
Near Future}
I already pointed out that the separation velocities of fragments that
are released from the nuclei of sungrazing comets at close proximity of
perihelion must be extremely low.  The widely ranging orbital periods of
the fragments can actually be explained even if they~\mbox{separate}~at
(relative) rest.  This is so because of the conditions~at~the~instant of
separation.  Whereas the fragment and the remaining part of the nucleus
still share the same orbital velocity, the heliocentric distances of their
respective centers of mass slightly differ.  As a result, the fragment and
the rest of the nucleus, which itself becomes a fragment, are bound to end
up in different orbits, their periods depending on their dimensions.

From the expression for the orbital velocity of a body in the elliptic
motion one obtains for a fragment's orbital period, $P_{\rm frg}$, with
high accuracy:
\begin{equation}
P_{\rm frg} = P_{\rm par} \! \left( \!1 - \frac{2U}{r_{\rm frg}^2}
 P_{\rm par}^{\frac{2}{3}} \!\!\right)^{\!\!\!-\frac{3}{2}} \!\!,
\end{equation}
where $P_{\rm par}$ is the orbital period of the pre-split parent comet,
$r_{\rm frg}$ is the heliocentric distance at fragmentation, and $U$ is
the distance of the center of mass of the fragment from the center of
mass of the parent comet along the radius vector; $U$ is negative when
the fragment is nearer the Sun at the separation time and vice versa,
and $|U|$ can be viewed as a lower limit of the sum of effective
semidiameters of the fragment and the parent's remnant.  The periods
$P_{\rm frg}$ and $P_{\rm par}$ are to be expressed in years and
$r_{\rm frg}$ and $U$ in AU.

The variations in the fragment's orbital period with the three parameters
are as follows:
\begin{eqnarray}
\frac{\partial P_{\rm frg}}{\partial P_{\rm par}} & = & \left( \! \frac{P_{\rm
 frg}}{P_{\rm par}} \!\right)^{\!\!\frac{5}{3}} \!\!\!, \nonumber \\[0.2cm]
\frac{\partial P_{\rm frg}}{\partial U} & = & \frac{3}{r_{\rm frg}^2}
 P_{\rm frg}^{\frac{5}{3}} , \nonumber \\[0.2cm]
\frac{\partial P_{\rm frg}}{\partial r_{\rm frg}} & = & -\frac{6U}{r_{\rm
 frg}^3} P_{\rm frg}^{\frac{5}{3}} = -\frac{2U}{r_{\rm frg}}
 \frac{\partial P_{\rm frg}}{\partial U} \,.
\end{eqnarray}

The rates of variation in $P_{\rm frg}$ as a function of the three
parameters share two important fearures:\ they increase with increasing
$P_{\rm frg}$ and decreasing $r_{\rm frg}$.  An obvious inference
from these results is that when fragmentation takes place very close to
perihelion, fragments of Population~I objects should get scattered over
periods of time nearly three times as wide as fragments of Population~II
objects.  Accordingly, fragmentation at perihelion implies that ---
everything else being equal --- one is more likely to witness the
return to perihelion of a sungrazer of Population~II than Population~I.

Overall, for fragments of a sungrazing comet separating at close proximity
to perihelion, the effect on the orbital period by a shift in $U$ of several
kilometers could be dramatic, often reaching hundreds of years, as seen
from Table~7. In the light of these computations, the difference of some
300~years between the orbital periods of Fragments~A and D of the Great
September Comet of 1882 (see Table~6), derived by Kreutz (1891), is not
in the least surprising.

Equation (7) is applicable of course not only to a relationship of the
parent and a fragment, but between any two fragments as well.  Furthermore,
one can turn the problem around and ask what radial distance between the
centers of mass of two fragments is needed to account for their respective
times of appearance.  Applied to the pair of the Great September Comet
of 1882 and Ikeya-Seki in 1965, with the Chinese comet of 1138 as their
previous return to perihelion, one finds \mbox{$P_{\rm 1882} = 744.1$ yr}~and
\mbox{$P_{\rm 1965} = 827.2$ yr}, and the radial distance between their
centers of mass, $U(1882, 1965)$, then equals
\begin{equation}
U(1882,1965) = {\textstyle \frac{1}{2}} \, r_{\rm frg}^2 \!\left(\!
 P_{\rm 1882}^{-\frac{2}{3}} - P_{\rm 1965}^{-\frac{2}{3}} \!\right) =
 0.000415 \, r_{\rm frg}^2,
\end{equation}
where the units are those used in Equation~(7).  Expressing now
$U(1882,1965)$ in km and $r_{\rm frg}$ in units of the Sun's radius,
one has
\begin{equation}
U(1882,1965) = 1.34 \,r_{\rm frg}^2.
\end{equation}
An absolute minimum value of $U$ is obtained on the assumption that
the heliocentric distance at fragmentation equaled the perihelion
distance of the parent comet in 1138, \mbox{$r_{\rm frg} = q =
1.73\,R_\odot$} (Sekanina \& Kracht 2022).  This option gives
\begin{equation}
U_{\rm min}(1882,1965) = 4.01 \; {\rm km},
\end{equation}
a somewhat low estimate.  However, the data presented in Section~3.13 suggested
that the 1882 and 1965 sungrazers were likely to have separated from one
another near \mbox{$r_{\rm frg} \simeq 4.1\,R_\odot$} (equivalent
to a fragmentation time of $\sim$2.4~hr after perihelion), in which case the
effective distance between the centers of mass of the two sungrazers
comes out to be
\begin{equation}
U_{\rm eff}(1882,1965) \simeq 23 \; {\rm km},
\end{equation}
a more realistic value.  Even this is still a lower limit, because
the line connecting the centers of mass is assumed in Equation~(12)
to be aligned with the radius vector.

For a comet's constant initial orbital period, $P_{\rm par}$, Table~8
demonstrates the high degree of sensitivity of the orbital period of a
fragment on the heliocentric distance, $r_{\rm frg}$, of the location
at which it separated from the parent nucleus.  The dependence of
$P_{\rm frg}$ on the center-of-mass distance $U$ suggests for example
that, compared to $1.1\,R_\odot$, a breakup at $3.1\,R_\odot$ changes
the fragment's orbital period by an amount that in some cases is smaller
by more than one order of magnitude.

\begin{table}[t] 
\vspace{0.17cm}
\hspace{-0.18cm}
\centerline{
\scalebox{0.99}{
\includegraphics{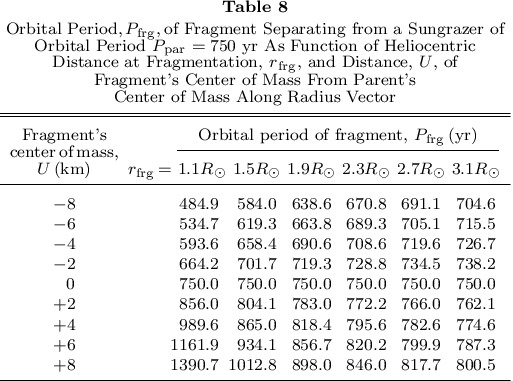}}}
\vspace{0.5cm}
\end{table}

The principal task of this section is to find out what~do the historical
records of bright sungrazers and the major effects of near-perihelion
fragmentation mean in terms of the prospects for the appearance of
one or more naked-eye Kreutz comets in the near future.  It turns out
that the times between appearances of suspected sungrazing comets,
which before 1882 were believed to indicate a single object's orbital
period (or a multiple thereof), are in fact determined by the {\it
differences\/} between the orbital periods of fragments whose origin
dates back to the perihelion passage of their parent.

In practice, the gained insight is somewhat impaired by
the existence of populations or subgroups, which had not been fully
acknowledged until the arrivals of comets Pereyra and Ikeya-Seki
in the 1960s.\footnote{Even though Kreutz always tested the suspected
sungrazers of the late 17th century and early 18th century to see
whether their orbits resembled closer the orbit of the 1843 or the
1882 sungrazer, he not once used the term subgroup.  Perhaps
justifiably, given that the number of sungrazers with orbits of good enough
quality known in his time was exactly three. (The orbit of Southern
Comet of 1887 was not then known well enough to be classified.)}  As
the populations are independent of one another and their number has
grown, the problem of dealing with prospects for future appearances of
bright sungrazers has been getting ever more intricate and convoluted.
In any case, to facilitate predictions, the problem has first to be
categorized by population.

Even then, the prognostication of future spectacular sungrazers is
risky and dependent on conditions such as (i)~the availability of a
feasible algorithm that governs the process of fragmentation in the
past and (ii)~a good chance that the algorithm will continue to govern
the process in the future.  This line of attack, the only one I know
of, still ignores the ever present stochasticity of the~process,
introducing uncertainty in proportion to the degree of its influence.
The method is very approximate anyway, because it does not distinguish
between the heliocentric and barycentric orbital periods (which,
strictly, should be taken at the time of fragmentation) and also
neglects effects of the planetary perturbations.

\begin{table}[t] 
\vspace{0.18cm}
\hspace{-0.21cm}
\centerline{
\scalebox{1}{
\includegraphics{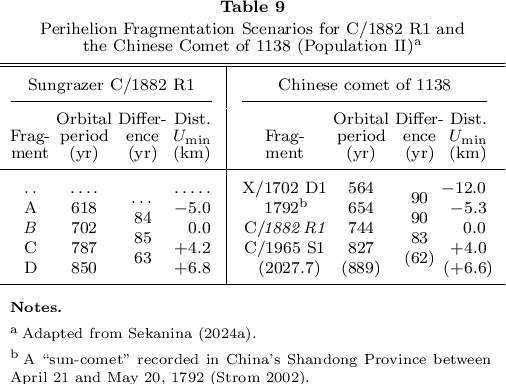}}}
\vspace{0.5cm}
\end{table}

Potentially helpful information on Population~II was offered by the
observed nuclear multiplicity of the 1882 sungrazer (Table~6).  Given
that this sungrazer and comet Ikeya-Seki were fragments of the same
object (Marsden 1967), equated with the Chinese comet~of~1138~(\mbox{Sekanina}
\& Kracht 2002), one could compare, side by side, fragmentation sequences
in comets of two consecutive generations --- the 1138 and 1882 sungrazers.
Now, if the 1882 comet was the most massive fragment of the 1138 comet
and nucleus~B the most massive fragment of the 1882 comet, it makes sense
to search for potential similarities in the relationship between the 1882
sungrazer and comet Ikeya-Seki --- the two successive fragments of the
1138 comet --- on the one hand and between nuclei~B and C --- the two
successive fragments of the 1882 comet --- on the other hand.  Table~9,
adapted from Sekanina (2024a), shows that this search revealed a remarkable
coincidence:\ the difference between the future (barycentric) orbital
periods of nuclei~B and C (85~yr) {\it equaled\/} almost exactly
the time between the arrivals of the 1882 and 1965 sungrazers (83~yr),
that is, the difference between the barycentric orbital periods of the
two comets.  Note that column~2 of the table is identical with the last
column of Table~6.  

As seen from Equations (8), the changes in the fragments' orbital periods,
$\Delta P_{\rm frg}$, depend on all three parameters, so that their small
variability in Table~9, while viewed positively, provides no path to easy
interpretation.  Instead, I used $\Delta P_{\rm frg}$ to compute the
center-of-mass distances $U_{\rm min}$ on the assumption that
fragmentation occurred at perihelion.  Although I showed above that
because of inertia this is not a good approximation, $U_{\rm min}$
could still be used to test the degree of similarity between the 1882
sungrazer and its parent.  While the sun-comet of 1792 is a weak point,
Table~9 suggests that the trends in $U_{\rm min}$ indeed are nearly identical,
from $-$5.0 to +4.2 for the 1882 sungrazers's nuclei A through C against
$-$5.3 to +4.0 for the parent's equivalent fragments.  The bonus is
that~$U_{\rm min}$~for nucleus~D~implies the arrival of an equivalent fragment
of the comet of 1138 {\it in the future\/}, thereby rendering the {\it
predictive capability\/}!~And Table~9~\mbox{delivers} an
undeniably~stunning~\mbox{prognosis}:~{\small \bf the~next~naked-eye sungrazer
of~\mbox{Population}~II is predicted to appear in the coming years}.  Year
2027 itself should not be taken verbatim, because the uncertainty is
at least several years either direction.  In this context, was~C/2024~S1
a precursor?  Given how rare the bright dwarf sungrazers of Population~II
are, it may very well have been!

\begin{table}[t] 
\vspace{0.18cm}
\hspace{-0.21cm}
\centerline{
\scalebox{1}{
\includegraphics{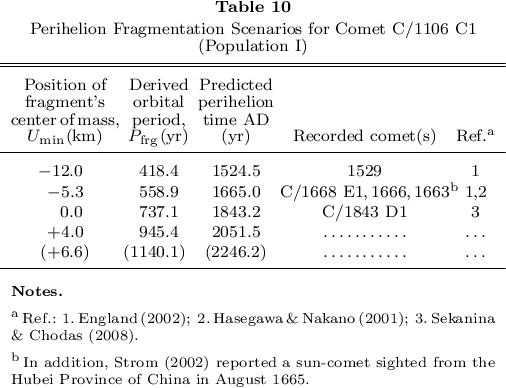}}}
\vspace{0.4cm}
\end{table}

I now proceed one step further and apply the same test to Population~I.
Here one is on extremely thin ice, because there is absolutely no
evidence for fragmentation of the Great March Comet of 1843 in close
proximity of perihelion.  What was then the chance that the 1843
sungrazer's parent, the Great Comet of 1106, {\it did\/} fragment near
its perihelion?  The only evidence that it was then losing mass has been
the existence of the SOHO stream of dwarf sungrazers, dominated by
Population~I and recently successfully modeled (Sekanina 2024b).
It should be pointed out that comet Pereyra, which appeared in 1963,
120~years after the Great March Comet of 1843, was not strictly a
member of Population~I (even though the orbits were similar), because
its long barycentric orbital period, exceeding 900~years, was
incompatible with the Great Comet of 1106 as its parent.

Unexpectedly, Table 10 offers {\small \bf another major surprise}:\
\mbox{$U_{\rm min} = -5.3$ km}, copied from Table~9, predicts that a
{\small \bf sibling comet of C/1843~D1 should have arrived} in 1665,
which differs by merely {\small \bf three years from} the time of
appearance of the celebrated {\small \bf comet C/1668~E1}, considered
by many in the early times to have been the 1843 sungrazer's previous
return to perihelion (Section~3.7).  Hasegawa \& Nakano (2001) listed
two other comets in the 1660s as further suspects, while Strom (2002)
found a sun-comet in Chinese records precisely in the predicted year
1665!  Secondary disruptions of the primary fragment at larger
heliocentric distances could account for a fairly tight swarm of
sungrazers, if the comets indeed were related.  The other negative value
of $U_{\rm min}$ predicted a sungrazer around the year 1524.  England's
(2002) possible candidate arrived five years later, but its ranking
was only~3.  Hasegawa \& Nakano (2001) offered no suspect within
50~years of the predicted time.

Next, I checked the prognosis for the positive values of
$U_{\rm min}$:\ one bright comet of Population~I is predicted to appear
near 2050 and another one in the mid-23rd century.  The first
of the two could be comparable in brightness to the 1668 comet.

The two predictions in the course of the next 30 or so years, an
exciting one for Population~II and less dramatic for Population~I,
are all that can be offered at this time.  Similar predictions for
Population~IIa and others are precluded by the insufficient database
available.{\vspace{0.1cm}}

Finally, two important, but highly speculative points.  The first is an
extension of the problem~of~a~\mbox{precursor}.\footnote{Not to be confused
with the precursors in the pedigree chart in Figure~3, which represented
an intermediate stage between the lobes and the first-generation fragments
of the various populations of the Kreutz system.{\vspace{-0.1cm}}}  In a
recent paper (Sekanina 2024a), the bright dwarf sungrazer C/2024~S1 (ATLAS),
unquestionably of Population~II, was compared with C/1945~X1 (du Toit).
This comparison assumed that du Toit was a member of the same population,
which was possible but by no means certain.  However, if it was, it makes
sense to ask whether du~Toit might have been a precursor to Ikeya-Seki,
at perihelion 20~years later, and comet ATLAS a precursor to the bright
sungrazer predicted to arrive in a few years.  If so, the arrival time
of ATLAS suggests --- when approximately scaled from du~Toit --- that
the predicted comet may not arrive until about the year 2039, intimating
the prognosis uncertainty of at least 12~years.  I see no lead to removing
the large error.

The second point concerns major fragments of Aristotle's comet that had
separated from it {\it before\/} the birth of the Kreutz system and survived
(in whatever size range) to this time.  I have in mind large chunks
from breakup events that occurred at close proximity of perihelion in
372~BC, less those at the previous perihelion in 1125~BC, and least
those at the perihelion in 1901~BC (Sekanina \& Kracht 2022).  The
computed barycentric orbital periods were 734.8~yr between 372~BC and
AD~363, 752.3~yr between 1125~BC and 372~BC, and 776.8~yr between
1901~BC and 1125~BC.  Reckoned from the present time, the intervals
of time to the three perihelia were 2396~yr, 3148~yr, and 3925~yr,
respectively.  Thus, a fragment separating in 372~BC would return to
perihelion in a few years from now, if its mean orbital period has been
just about 800~yr, surviving over three revolutions about the Sun; or if
it has been just about 600~yr, surviving over four revolutions. (I assume
that no sungrazer had an extreme mean orbital period of 1200~yr, surviving
over two revolutions.)  Similarly, a fragment separating in 1125~BC would
return to perihelion in a few years from now, if its mean orbital period
has been 790~yr, surviving four revolutions; or if it has been 630~yr,
surviving five revolutions; etc.  However, a survival time of more than
2--3~revolutions is unlikely, at least in the context of the contact-binary
model, which assumes that most mass of the progenitor was in the two
lobes, which were still essentially intact when they separated from
one another to give birth to the Kreutz system.  The fragments that
separated from the progenitor earlier should have been smaller (perhaps
a few kilometers across at most) and the best chance for surviving up to
now have had the ones that detached most recently --- the 372~BC fragments.

Before I deal with some of those fragments, I remark that a computed
orbit of Aristotle's comet at its 372~BC return had a perihelion distance
of 0.0068~AU and future orbital period of 735~yr, so that \mbox{$U_{\rm
min} = -12.0$ km}, $-$5.3~km, +4.0~km, and +6.6~km (Table~9) lead,
respectively, to \mbox{$P_{\rm frg} = 506$ yr}, 616~yr, 853~yr, and
947~yr.  The candidate sungrazers should have appeared in AD~135, 245,
482, and 576.  It is encouraging to see that among the Kreutz suspects
in Hasegawa \& Nakano's (2001) list is a comet in AD~133 and another one
in 245.  Then they, as well as \mbox{England} (2002), show the comet of
467, 15~yr too early to fit \mbox{$U_{\rm min} = +4.0$ km}.  According to
Mart\'{\i}nez et al.\ (2022) this comet moved in a Kreutz-like orbit and,
together with the comet of 423, was at one point judged a possible
5th-century appearance of the Great Comet of 1106 (Sekanina \& Chodas
2007).  Nearest to AD~576 in Hasegawa \& Nakano's list is a comet in
607, an unlikely match, but England has a Kreutz suspect in 582.\,

A fragment with a mean orbital period of 800~yr would return in AD~429
and the comet that fits fairly well this scenario is of course the comet
of 423, the other 5th-century suspect for the previous appearance of the
1106 comet.  Unfortunately, 800~yr is more than 50~yr away from the
nearest likely orbital period of a fragment separating from Aristotle's
comet at the 372~BC perihelion (\mbox{$U_{\rm min} = +4.0$ km}).  More
hopeful is the orbital period of 600~yr, as \mbox{$U_{\rm min} = -5.3$ km}
fits not only the comet of 245, but predicts the following return in the
year 861.  Both Hasegawa \& Nakano and England have a Kreutz suspect in
AD~852, England another one in 867.  While it may sound disappointing
that there is no suspect at all in the 15th century, the sungrazer may
have by then crumbled to display instead a SOHO-like stream of dwarf
comets.  This is perhaps a more attractive explanation than the
traditional seasonal effect of missed sungrazers.

One characteristic property of the fragments released from Aristotle's
comet at perihelion is that their longitude of the ascending node is near
that of the progenitor, so that in Figure~1 their debris should appear
as part of Population~Ia, or very close to it, and may be confused with
the debris from the neck of the binary.  Or, perhaps even more probably,
if the neck detached from Lobe~I {\it after\/} the lobes separated from
each other, the neck debris may be concentrated in Population~I$_0$, whose
origin is otherwise somewhat mysterious, and Population~Ia contains
exclusively the debris from the progenitor's very old fragments, such
as the comets of 467 and 852.  Hence, one cannot rule out a contamination
of the histogram in Figure~1 by a debris of Aristotle's comet that strictly
does not belong to the Kreutz system.  Rather intriguing!

\section{Conclusions} 
In early days, recurring appearances of bright comets near the Sun over
the periods of time amounting to  from tens to 200 or so years were usually
thought to be returns of a single object, the observed times providing
information on its orbital period.  Astrometry in those days was so poor
that no orbit of adequate quality could be computed to actually determine
the period.  When three or more comets were involved, the treatment
of the problem was all arithmetic and no astronomy.  If the starting
premise of a single body was incorrect (which,~\mbox{except}~for Halley's
comet, almost universally was the case),~the~solution usually implied
unrealistically short periods,~the shorter the larger number of comet
appearances was~to~be accommodated.  As a result, the solution was open
to the compelling objection of an unacceptably large number of returns
at which the comet was not observed.

An appalling example was the widespread belief --- after the appearance
of sungrazer C/1843~D1 --- that this was a return of comet C/1668~E1,
thus implying an orbital period of 175~years.  However, the new sungrazer
was also thought by many to be a return of the Great Comet of 1106.
Could both identities be true simultaneously?  The problem was
that 1843--1106 = 737~years and 737/175 = 4.21, far from an integer.
This started the search for an orbital period that should make the
comet return in 1106, 1668, as well as 1843.  That period, proposed
by Laugier \& Mauvais (1843b), was equal to 35.1~years:\ the comet
would have made 16~revolutions between 1106 and 1668 and 5~revolutions
between 1668 and 1843.  But it also should have appeared in 1141,\,1176,
1211,$\:$1246,$\:$1281,$\:$1316,$\:$1351,$\:$1387,$\:$1422,$\:$1457,$\:$1492,
1527, 1562,$\:$1597,$\:$1633,$\:$1703,$\:$1738,$\:$1773,$\:$and$\:$1808,
not~to~men- tion the appearances prior to 1106.  Inspection~of~Hase\-gawa
\& Nakano's (2001) and England's (2002) lists of historical Kreutz sungrazer
candidates and Strom's (2002) table of sun-comets shows that not one
of the suspects was observed in any of the above 16~particular years
between 1141 and 1703.  Likewise, no Kreutz sungrazers and no sun-comets
were reported in 1738, 1773, or 1808.  This failure contributed to the
rejection of the hypothesis of a single returning sungrazing
object.{\vspace{0.02cm}}

The point that I have tried to make in this paper is that even though
scientific interest in the sungrazers began at least three centuries
ago, it was less than 150~years ago that a watershed in understanding
the nature of sungrazers and, to a degree, comets in general took
place following the dramatic show of the fragmenting nucleus of the
Great September Comet of 1882 shortly after its perihelion passage.
It was this event that not only discredited the hypothesis of a single
returning object, with which comet astronomers were obsessed before
1882, but, thanks to Kreutz's (1891) commendable computations, it
finally became indisputable that the orbital periods of the sungrazers
were in the general range of 500--1000~yr, whereas the time intervals
shorter than 200~yr, taken mistakenly for the orbital periods before
1882, were in fact {\it differences\/} between the orbital periods
of the individual nuclear fragments.  Kreutz was also able to link
the preperihelion observations of the original nucleus of the 1882
sungrazer with the post-perihelion observations of the nuclear
fragments to show that there was no measurable effect of the solar
corona on the orbital motion.{\vspace{0.02cm}}

Whereas Kreutz could surely claim the lion's share of the credit for
the progress in understanding the motions of the sungrazing comets
in the late 19th century, Hall (1883) was the first to propose that
Aristotle's comet, the Great Comet of 1106, and the Great March Comet
of 1843 were three appearances of the same object, pointing out that
there was no record of the return in 368.\footnote{Maxwell Hall
(1845--1920), not to be confused with his illustrious American
contemporary Asaph Hall (1829--1907), the discoverer of Phobos and
Deimos, or Asaph Hall,~Jr.\ (1859--1930), was a British astronomer
and meteorologist, who in 1872 settled in Jamaica, where he erected
the Kempshot Observatory at Montego Bay.  For obituary, see Mon.\
Not.\ Roy.\ Astron.\ Soc., 81, 259 (1921).}  On the other hand, Kreutz
(1888, 1901) believed that the 1106 comet returned as the sungrazer of
1882, while Hall had very little to say about the history of that
object.{\vspace{0.02cm}}

The next major step forward was the recognition of two subgroups, or
populations, of sungrazers thanks to comets Pereyra and Ikeya-Seki,
which arrived in 1963 and 1965, respectively, and increased the number
of sungrazers with well-determined orbits from three to five.  The
ramifications were investigated by Marsden (1967), whose most
significant contribution was virtual proof that Ikeya-Seki and the
Great September Comet of 1882, the two Population~II members among
these five sungrazers, constituted a single object on the way to the
previous perihelion in the early 12th century.  This remarkable
condition was satisfied over a broad range of assumed perihelion
times:\ for 1106 (Marsden 1967) and even a little better for 1138
(Sekanina \& Kracht 2022).

On the other hand, the large differences between Populations~I and II
in the angular elements and perihelion distance could not be explained
by fragmentation at perihelion, and Marsden (1967) realized that this was
a major problem.  The things got worse, when two new bright sungrazers,
appearing in 1970 and 2011, respectively, did not belong to either of the
two populations and further extended the orbital range of the Kreutz system.
In the meantime, its perception has been changing immensely in the era of
coronagraphic discoveries of faint sungrazers from space, especially the
Solar and Heliospheric Observatory.  The episodic distribution of dwarf
sungrazers, many arriving in close pairs or swarms, has been a product
of nontidal fragmentation proceeding along the entire orbit and in a
cascading fashion (Sekanina 2000, 2002).  As an ultimate outcome of
this process, enormous numbers of fragments with sizes down to less
than 10~meters in diameter disintegrate and sublimate away completely
on their terminal approach to perihelion.  A constant flow of vast
amounts of new information in the past decades has necessitated a
thorough revision and update of the model for the Kreutz system.
The pyramidal contact-binary model (Sekanina~2021, 2022a, 2022b,
2022c, 2023, 2024a, 2024b, Sekanina \& Kracht 2022) has been a
response to satisfy these needs.

In the context of this model and as a test of some of its potential
implications, I present in this paper specific predictions of two
bright, naked-eye Kreutz sungrazers to appear over the next thirty
years or so; the prognosis is based on certain patterns believed to
be detected in the past and the uncertainty is up to approximately
ten or so years.  The exciting chance is the predicted arrival of a
major member of Population~II, possibly comparable to Ikeya-Seki
in terms of {\it intrinsic\/} brightness, over the next five years,
less probably at some time before 2040.  The other prognosis concerns
a major sungrazer~of Population~I, which, predicted to appear about 2050,
could be comparable to C/1668~E1 in intrinsic brightness.{\vspace{-0.09cm}}

%
%
%
%
%
\begin{center}
{\footnotesize REFERENCES}
\end{center}
\vspace{-0.5cm}
\begin{description}
{\footnotesize
\item[\hspace{-0.3cm}]
Arago, F.\ 1843, Compt.\ Rend., 16, 605, 639, 718
\\[-0.57cm]
\item[\hspace{-0.3cm}]
Barnard, E.\ E.\ 1883, Astron.\ Nachr., 104, 267
\\[-0.57cm]
\item[\hspace{-0.3cm}]
Bernardi\`eres, O.\ de 1882, Compt.\ Rend., 95, 823
\\[-0.57cm]
\item[\hspace{-0.3cm}]
Boguslawski, M.\ von 1845, Astron.\ Nachr., 23, 269
\\[-0.57cm]
\item[\hspace{-0.3cm}]
Burckhardt, J.\ C.\ 1817, Conn.\ Temps 1817, 278
\\[-0.57cm]
\item[\hspace{-0.3cm}]
Capocci, E.\ 1843, Astron.\ Nachr., 21, 155; also:\ Mon.\ Not.\ Roy.{\linebreak}
 {\hspace*{-0.6cm}}Astron.\ Soc., 5, 298
\\[-0.57cm]
\item[\hspace{-0.3cm}]
Cassini, G.\ D.\ 1668, Phil.\ Trans.\ Roy.\ Soc., 3 (35), 683
\\[-0.57cm]
\item[\hspace{-0.3cm}]
Cassini, G.\ D.\ 1702, M\'em.\ Paris Acad.\ Sci.\ 1702, 101
\\[-0.57cm]
\item[\hspace{-0.3cm}]
Clausen, Th.\ 1843, Astron.\ Nachr., 21, 73
\\[-0.57cm]
\item[\hspace{-0.3cm}]
Cooper,~E.\,J.\,1852,~Cometic~Orbits.~Dublin:\,A.\,Thom~Publ.,\,196pp
\\[-0.57cm]
\item[\hspace{-0.3cm}]
Encke, J.\ F.\ 1836, Astron.\ Nachr., 13, 263
\\[-0.57cm]
\item[\hspace{-0.3cm}]
England, K.\ J.\ 2002, J.\ Brit.\ Astron.\ Assoc., 112, 13
\\[-0.57cm]
\item[\hspace{-0.3cm}]
Frisby, E.\ 1883, Astron.\ Nachr., 104, 159 (erratum:\ 283); also:\
 Nat.,{\linebreak}
 {\hspace*{-0.6cm}}27, 226
\\[-0.57cm]
\item[\hspace{-0.3cm}]
Galle,\,J.\,G.\ 1894, Cometenbahnen. Leipzig:\,Verlag W.\,Engelmann,{\linebreak}
 {\hspace*{-0.6cm}}315pp
\\[-0.57cm]
\item[\hspace{-0.3cm}]
Gill, D.\ 1882, Mon.\ Not.\ Roy.\ Astron.\ Soc., 43, 19
\\[-0.57cm]
\item[\hspace{-0.3cm}]
Gill, D.\ 1883, Mon.\ Bot.\ Roy.\ Astron.\ Soc., 43, 319
\\[-0.57cm]
\item[\hspace{-0.3cm}]
Gill, D.\ 1911, Ann.\ Cape Obs., 2, Pt.\ 1, 3
\\[-0.57cm]
\item[\hspace{-0.3cm}]
Golvers, N.\ 2014, Almagest, 5 (1), 32
\\[-0.57cm]
\item[\hspace{-0.3cm}]
Golvers, N.\ 2025, Almagest, 9 (2), 88
\\[-0.55cm]
\item[\hspace{-0.3cm}]
Hall, M.\ 1883, Obs., 6, 233
\\[-0.39cm]
\item[\hspace{-0.3cm}]
Hasegawa, I.\ 1980, Vistas Astron., 24, 59
%
\item[\hspace{-0.3cm}]
Hasegawa, I., \& Nakano, S.\ 2001, Publ.\ Astron.\ Soc.\ Japan, 53, 931
\\[-0.57cm]
\item[\hspace{-0.3cm}]
Henderson, T.\ 1843, Astron.\ Nachr., 20, 333;
also:\ Mon.\ Not.\ Roy.{\linebreak}
 {\hspace*{-0.6cm}}Astron.\,Soc., 5, 267
\\[-0.57cm]
\item[\hspace{-0.3cm}]
Hoek, M.\ 1865a, Mon.\ Not.\ Roy.\ Astron.\ Soc., 25, 243
\\[-0.57cm]
\item[\hspace{-0.3cm}]
Hoek, M.\ 1865b, Mon.\ Not.\ Roy.\ Astron.\ Soc., 26, 1
\\[-0.57cm]
\item[\hspace{-0.3cm}]
Hoek, M.\ 1866, Mon.\ Not.\ Roy.\ Astron.\ Soc., 26, 204
\\[-0.57cm]
\item[\hspace{-0.3cm}]
Holetschek, J.\ 1892, Astron.\ Nachr., 129, 323
\\[-0.57cm]
\item[\hspace{-0.3cm}]
Hubbard, J.\ S.\ 1851, Astron.\ J., 2, 57
\\[-0.57cm]
\item[\hspace{-0.3cm}]
Hubbard, J.\ S.\ 1852, Astron.\ J., 2, 153
\\[-0.57cm]
\item[\hspace{-0.3cm}]
Kalinicheva, O.\ V.\ 2017, Solar Syst.\ Res., 51, 221 
\\[-0.57cm]
\item[\hspace{-0.3cm}]
Kendall, E.\ O.\ 1843, Astron.\ Nachr., 20, 387; also:\ Mon.\ Not.\ Roy.{\linebreak}
 {\hspace*{-0.6cm}}Astron.\ Soc., 5, 304
\\[-0.57cm]
\item[\hspace{-0.3cm}]
Kirkwood, D.\ 1880, Obs., 3, 590
\\[-0.57cm]
\item[\hspace{-0.3cm}]
Kreutz, H.\ 1888, Publ.\ Sternw.\ Kiel, 3
\\[-0.57cm]
\item[\hspace{-0.3cm}]
Kreutz, H.\ 1891, Publ.\ Sternw.\ Kiel, 6
\\[-0.57cm]
%
%
\item[\hspace{-0.3cm}]
Kreutz, H.\ 1901, Astron.\ Abhandl., 1, 1
\\[-0.57cm]
\item[\hspace{-0.3cm}]
Kronk, G.\ W.\ 1999, Cometography, Volume 1:\ Ancient--1799.{\linebreak}
 {\hspace*{-0.6cm}}Cambridge, UK:\ University Press, 580pp
\\[-0.57cm]
\item[\hspace{-0.3cm}]
Laugier, M., \& Mauvais, V.\ 1843a, Compt.\ Rend., 16, 781
\\[-0.57cm]
\item[\hspace{-0.3cm}]
Laugier, M., \& Mauvais, V.\ 1843b, Compt.\ Rend., 16, 919
\\[-0.57cm]
\item[\hspace{-0.3cm}]
Lynn, W.\ T.\ 1882, Obs., 5, 329
\\[-0.57cm]
\item[\hspace{-0.3cm}]
Marsden, B.\ G.\ 1967, Astron.\ J., 72, 1170
\\[-0.57cm]
\item[\hspace{-0.3cm}]
Marsden, B.\ G.\ 1989, Astron.\ J., 98, 2306
\\[-0.57cm]
\item[\hspace{-0.3cm}]
Marsden, B.\ G.\ 2005, Annu.\ Rev.\ Astron.\ Astrophys., 43, 75
\\[-0.57cm]
\item[\hspace{-0.3cm}]
Marsden, B.\ G., \& Williams, G.\ V.\ 2008, Catalogue of Cometary{\linebreak}
 {\hspace*{-0.6cm}}Orbits 2008 (17th ed.)  Cambridge, MA:\ IAU Central Bureau
 for{\linebreak}
 {\hspace*{-0.6cm}}Astronomical Telegrams and Minor Planet Center, 195pp
\\[-0.57cm]
\item[\hspace{-0.3cm}]
Marsden,\,B.\,G., Sekanina,\,Z.,\,\&\,Everhart,\,E.\ 1978, Astron.\,J.,\,83,\,64
\\[-0.57cm]
%
%
\item[\hspace{-0.3cm}]
Mart\'{\i}nez, M.\,J., Marco, F.\,J., Sicoli, P., \& Gorelli, R.\
 2022,~Icarus,{\linebreak}
 {\hspace*{-0.6cm}}384, 115112
\\[-0.57cm]
\item[\hspace{-0.3cm}]
Maury, M.\ F.\ 1846, Astron.\ Nachr., 24, 135
\\[-0.57cm]
\item[\hspace{-0.3cm}]
Meyer, M.\ W.\ 1880, Astron.\ Nachr., 97, 185
\\[-0.57cm]
\item[\hspace{-0.3cm}]
Meyer, M.\ W.\ 1882, Astron.\ Nachr., 102, 83
\\[-0.57cm]
\item[\hspace{-0.3cm}]
Millon, D.\ 1966, Strol.\ Astron., 19, 206
\\[-0.57cm]
\item[\hspace{-0.3cm}]
Nicolai, B.\ 1843, Astron.\ Nachr., 20, 349
\\[-0.57cm]
\item[\hspace{-0.3cm}]
Oppolzer, Th.\ von 1880, Astron.\ Nachr., 97, 225
\\[-0.57cm]
\item[\hspace{-0.3cm}]
Pech\"{u}le, C.\ F.\ 1868, Astron.\ Nachr., 72, 235
\\[-0.57cm]
\item[\hspace{-0.3cm}]
Penrose, F.\ C.\ 1882, Mon.\ Not.\ Roy.\ Astron.\ Soc., 43, 25
\\[-0.57cm]
\item[\hspace{-0.3cm}]
Pingr\'e,\,A.\,G.\,1783,$\:$Com\'etographie$\:$ou\,Trait\'e$\:$historique$\:$et$\:$th\'eorique{\linebreak}
 {\hspace*{-0.6cm}}des com\`etes. Tome Premier.  Paris:\ L'Imprimerie Royale,
 630pp
\\[-0.57cm]
\item[\hspace{-0.3cm}]
Pingr\'e,\,A.\,G.\,1784,$\:$Com\'etographie$\:$ou\,Trait\'e$\:$historique$\:$et$\:$th\'eorique{\linebreak}
 {\hspace*{-0.6cm}}des com\`etes. Tome Second.  Paris:\ L'Imprimerie Royale,
 518pp
\\[-0.57cm]
\item[\hspace{-0.3cm}]
Plantamour, E.\ 1846, M\'em.\ Soc.\ Phys.\ Hist.\ Nat.\ Gen\`eve, 11, 42
\\[-0.57cm]
\item[\hspace{-0.3cm}]
Plummer, W.\ E.\ 1892, Obs., 15, 308
\\[-0.57cm]
\item[\hspace{-0.3cm}]
Rebeur-Paschwitz, E.\ von 1884, Astron.\ Nachr., 107, 383
\\[-0.57cm]
\item[\hspace{-0.3cm}]
Roche, \'E.\ 1849, M\'em.\ Sec.\ Sci.\ Acad.\ Montpellier, 1, 243
\\[-0.57cm]
\item[\hspace{-0.3cm}]
Roche, \'E.\ 1850, M\'em.\ Sec.\ Sci.\ Acad.\ Montpellier, 1, 333
\\[-0.57cm]
\item[\hspace{-0.3cm}]
Roche, \'E.\ 1851, M\'em.\ Sec.\ Sci.\ Acad.\ Montpellier, 2, 21
\\[-0.57cm]
\item[\hspace{-0.3cm}]
Rolfe, J.\ C.\ 1940, The Roman History of Ammianus Marcellinus,{\linebreak}
 {\hspace*{-0.6cm}}Book XXV. {\tt
 https://penelope.uchicago.edu/Thayer/E/Roman/}{\linebreak}
 {\hspace*{-0.6cm}}{\tt Texts/Ammian/25$^*\!\!$.html}
\\[-0.57cm]
\item[\hspace{-0.3cm}]
Schumacher, H.\ C.\ 1843, Astron.\ Nachr., 20, 397
\\[-0.57cm]
\item[\hspace{-0.3cm}]
Seargent, D.\ 2009, The Greatest Comets in History:\ Broom Stars{\linebreak}
 {\hspace*{-0.6cm}}and Celestial Scimitars. New York:\ Springer
 Science\,+\,Business{\linebreak}
 {\hspace*{-0.6cm}}Media, LLC, 260pp
\\[-0.57cm]
%
%
%
\item[\hspace{-0.3cm}]
Sekanina, Z.\ 1980, in Comets, L.\ L.\ Wilkening, ed.\ Tucson:\
 Univer-{\linebreak}
 {\hspace*{-0.6cm}}sity of Arizona, 251
\\[-0.57cm]
\item[\hspace{-0.3cm}]
Sekanina, Z.\ 2000, Astrophys.\ J., 542, L147
\\[-0.57cm]
\item[\hspace{-0.3cm}]
Sekanina, Z.\ 2002, Astrophys.\ J., 566, 577
\\[-0.57cm]
\item[\hspace{-0.3cm}]
Sekanina, Z.\ 2021, eprint arXiv:2109.01297
\\[-0.57cm]
\item[\hspace{-0.3cm}]
Sekanina, Z.\ 2022a, eprint arXiv:2202.01164
\\[-0.57cm]
\item[\hspace{-0.3cm}]
Sekanina, Z.\ 2022b, eprint arXiv:2211.03271
\\[-0.57cm]
\item[\hspace{-0.3cm}]
Sekanina, Z.\ 2022c, eprint arXiv:2212.11919
\\[-0.57cm]
\item[\hspace{-0.3cm}]
Sekanina, Z.\ 2023, eprint arXiv:2305.08792
\\[-0.57cm]
\item[\hspace{-0.3cm}]
Sekanina, Z.\ 2024a, eprint arXiv:2411.12941
\\[-0.57cm]
\item[\hspace{-0.3cm}]
Sekanina, Z.\ 2024b, eprint arXiv:2404.00887
\\[-0.57cm]
\item[\hspace{-0.3cm}]
Sekanina, Z., \& Chodas, P.\ W.\ 2004, Astrophys.\ J., 607, 620
\\[-0.57cm]
\item[\hspace{-0.3cm}]
Sekanina, Z., \& Chodas, P.\ W.\ 2007, Astrophys.\ J., 663, 657
\\[-0.57cm]
\item[\hspace{-0.3cm}]
Sekanina, Z., \& Chodas, P.\ W.\ 2008, Astrophys.\ J., 687, 1415
\\[-0.57cm]
\item[\hspace{-0.3cm}]
Sekanina, Z., \& Kracht, R.\ 2015, Astrophys.\ J., 801, 135
\\[-0.57cm]
\item[\hspace{-0.3cm}]
Sekanina, Z., \& Kracht, R.\ 2022, eprint arXiv:2206.10827
\\[-0.57cm]
\item[\hspace{-0.3cm}]
Sekanina, Z., Chodas, P.\ W., \& Yeomans, D.\ K.\ 1998, Plan.\ Space{\linebreak}
 {\hspace*{-0.6cm}}Sci., 46, 21
\\[-0.57cm]
\item[\hspace{-0.3cm}]
Strom, R.\ 2002, Astron.\ Astrophys., 387, L17
\\[-0.57cm]
\item[\hspace{-0.3cm}]
Thollon, L., \& Gouy, L.\ G.\ 1882, Compt.\ Rend., 95, 555
\\[-0.57cm]
\item[\hspace{-0.3cm}]
Valz, B.\ 1843, Compt.\ Rend., 16, 925
\\[-0.57cm]
\item[\hspace{-0.3cm}]
Vogel, E.\ 1852a, Mon.\ Not.\ Roy.\ Astron.\ Soc., 12, 206
\\[-0.57cm]
\item[\hspace{-0.3cm}]
Vogel, E.\ 1852b, Astron.\ Nachr., 34, 387
\\[-0.57cm]
\item[\hspace{-0.3cm}]
Weiss, E.\ 1880, Astron.\ Nachr., 97, 61
\\[-0.65cm]
\item[\hspace{-0.3cm}]
Wilson, H.\ C.\ 1883, Publ.\ Cincinnati Obs., 7, 62}
%
%
\vspace{-0.43cm}
\end{description}
\end{document}